\documentclass[a4paper,11pt]{report}

\usepackage{graphicx} 
\usepackage{setspace}
\usepackage[numbers,sort&compress]{natbib}
\makeatletter
    \def\thebibliography#1{\chapter*{References\@mkboth
      {REFERENCES}{REFERENCES}}\list
      {[\arabic{enumi}]}{\settowidth\labelwidth{[#1]}\leftmargin\labelwidth
	\advance\leftmargin\labelsep
	\usecounter{enumi}}
	\def\newblock{\hskip .11em plus .33em minus .07em}
	\sloppy\clubpenalty4000\widowpenalty4000
	\sfcode`\.=1000\relax}
    \makeatother

\begin{document}

\pagestyle{empty}
\begin{center}

\Huge{Numerical Green's Function Modeling of One-Dimensional Quantum Transport} \vspace{3 cm}

\LARGE{Raphael Chayim Rosen} \\ \vspace{1.5 cm}  \Large{Wolfson College} \\ \Large{University of Cambridge} 
\vspace{1 cm}

\includegraphics[scale=0.35]{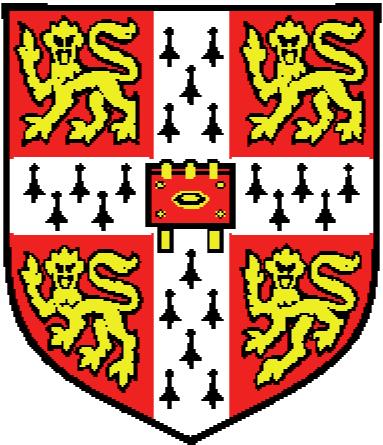}

\vspace{1 cm}
\normalsize{A dissertation submitted for the degree of \\ Master of Philosophy \\ \vspace{1cm} June 2007}

\end{center}

\clearpage
\pagestyle{plain}
\pagenumbering{roman}

\chapter*{Abstract}

Since the initial development of one-dimensional electron gases (1DEG) two decades ago, there has been intense interest in both the fundamental physics and the potential applications---including quantum computation---of these quantum transport systems. While experimental measurements of 1DEGs reveal the conductance through a system, they do not probe critical  other aspects of the underlying physics, including energy eigenstate distribution, magnetic field effects, and band structure. These are better accessed by theoretical modeling, especially modeling of the energy and wavefunction distribution across a system: the local density of states (DOS).

In this thesis, a numerical Green's function model of the local DOS in a 1DEG has been developed and implemented. The model uses an iterative method in a discrete lattice to calculate Green's functions by vertical slice across a 1DEG. The numerical model is adaptable to arbitrary surface gate geometry and arbitrary finite magnetic field conditions. When compared with exact analytical results for the local DOS, waveband structure, and real band structure, the model returned very accurate results. In zero magnetic field, the local DOS plots from the model behaved as anticipated by theory; under a finite magnetic field, depopulation and waveband separation were present in the model, also, precisely as was expected. The model was also used to investigate imaginary band structure and gave interesting results warranting further investigation. A second numerical model was also developed that measured the transmission and reflection coefficients through the quantum system based on the Landauer-B\"{u}ttiker formalism. The combination of the local DOS model with the transmission coefficients model was applied to two current research topics: antidot behavior and zero-dimensional to one-dimensional tunneling. These models can be further applied to investigate a wide range of quantum transport phenomena.

\chapter*{Declaration}

This dissertation is the result of my own work and, unless otherwise
stated, contains nothing which is the outcome of work done in
collaboration. No part of this thesis has already been, or is
currently being, submitted for any other qualification than the degree
of Master of Philosophy at the University of Cambridge. As mandated by the special regulations of the M.Phil. examination, this thesis contains fewer than 15,000 words.

\vspace{.4 cm}
Raphael C. Rosen

June 2007

\chapter*{Acknowledgments}

 \textit{``A teacher affects eternity; he can never tell where his influence stops."}

-Henry Adams
\vspace{1.6 cm}

Thank you to my advisor, Dr. C. H. W. Barnes for your humour, ready willingness to assist, and
breadth of knowledge.

To Prof. Sir M. Pepper, thank you for the privilege of working in your group, and for frequently checking in. I am grateful to the many researchers in the Semiconductor Physics Group who assisted me throughout my work. Many thanks to Samir Rihani, Raj Patel, Stephen Sarkozy, Mamta Thangaraj, Sieglinde Pfaendler, Dr. Frank Lee, and Adam Thorn. 

I am indebted to Emma Faid, Fiona Winter, and Alison Dann for their help throughout the year.

Many thanks to those physicists not in condensed matter who read drafts of this thesis and made sure it was intelligible to those not in the immediate field: Peter Graham, Lucas Laursen, and Mordecai Rosen. 

I would like to acknowledge the Herschede Engineering Award Committee, the Balfour Fellowship Committee, and the Harvard Club of the United Kingdom Scholarship Committee for their generous financial support.

To Jeanette LGW, Zaydie, Ema, Abba, Gavri, Jesse, Michael, Debbie, Eytan, and Danya, thank you for all your love and support.

\textbf{This thesis is lovingly dedicated to my grandmother, Roslyn Brickman, of blessed memory, whose zeal for learning and teaching has inspired me throughout life.}

\pagestyle{headings}
\setcounter{tocdepth}{5}
\tableofcontents

\listoffigures

\clearpage
\pagenumbering{arabic}
\doublespacing
\chapter{Introduction}

The confinement of electron motion to a single dimension, an experimental achievement that opened new realms of physics, has garnered great interest since its development two decades ago \cite{Eugster}. As the limiting case in which current can be carried, one dimensional systems not only present fundamental physics challenges but also portend numerous opportunities for practical application \cite{Meyer}. Most notably, they show potential to serve as the backbone of a future quantum-information-computation system \cite{Berggren}. 

\begin{figure}
\centering
\includegraphics[width=0.9\textwidth, viewport= 110 220  350 370]{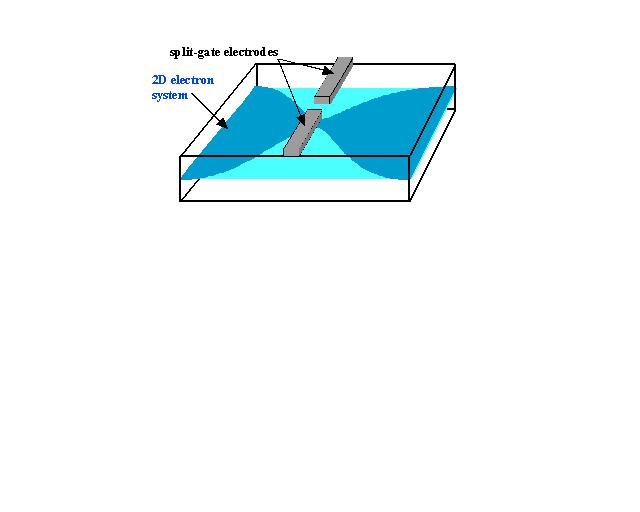}
\caption[Split Gate Device]{\textit {Split Gate Device. The metal gates sitting atop the semiconductor heterostructure deplete the region underneath, leaving only a tiny channel in the middle in which electrons can be present: a 1D channel}}\label{fig:splitgate.jpg}
\end{figure}

A one-dimensional electron system is obtained by modifying a two-dimensional electron gas (2DEG). A 2DEG is created within a semiconductor heterostructure, a stack of a few different semiconductors that takes advantage of band structure to achieve only one allowable energy level in the z-direction. Most commonly the one-dimensional modification of a 2DEG is generated by the use of the easily-adaptable split-gate device \cite{Eugster, Berggren}. Pioneered in the 1980's, a split-gate is a strip of metal with a narrow slit across its width that sits atop the heterostructure (see Figure \ref{fig:splitgate.jpg})  \cite{Thornton}. By applying a voltage to the gate, the heterostructure beneath it is depleted, leaving only a tiny channel through which the electrons can move: that is, the area underneath the slit. Since the width of this channel is roughly equal to the electron's wavelength, a one-dimensional electron gas (1DEG) is created. 

When van Wees \textit{et al}  \cite{vanWees} and Wharam \textit{et al} \cite{Wharam} independently developed the first one-dimensional systems in 1988, their demonstration of quantized conductance resolved a three decades-old theoretical debate (see \cite{Szafer}) and sparked numerous novel investigations. Resulting studies in 1D systems included investigations of the effects of high magnetic field and Zeeman splitting, of magnetic depopulation, of electric depopulation, and of tunneling between 1D channels \cite{Patel, Newson, Kouwen, Eugster}. These were followed by analyses of the behavior and number of occupied subbands and of the possible subband energies in 1D systems \cite{Kardynal, Macks}. Inquiries into interference effects have been carried out extensively for the last 20 years. 

Yet, the physics underlying these 1DEG systems remains incompletely understood. \cite{Beenakker, Graham}. Curiosity about 1DEGs has only accelerated in recent years, fueled by discoveries such as spontaneous spin splitting (the 0.7 structure) \cite{Thomas}. Further comprehension of experimental 1D systems promises to be advanced by a deeper theoretical grasp of 1DEGs underlying phenomena \cite{Bahder}.

What are these phenomena? They include the conductance, the capacitance and the density of states \cite{Kim, Bahder}. It is the density of states (DOS), in particular, that plays a fundamental role in understanding 1D systems \cite{Anwar}. A measure of the number of states available in a given energy range per unit length (or area), the DOS can provide details about a system's wavefunctions, resonant states, thermodynamics, scattering amplitudes and transmission probabilities \cite{Bruno, EconBook}. 

Given its centrality to quantized transport systems, it is not surprising that much work has been carried out on the DOS in semiconductor heterostructures. The total (bulk) DOS of a system has been investigated under periodic potentials \cite {Kramer, RBSA, RBSB}, under modulated magnetic fields \cite{RBSC, RBSD}, and in relation to localization length \cite{Nikolic, Half}. The effects of spin \cite{Wang} and non-resonant laser light \cite{Enders} on the DOS have also been pursued. The two-dimensional DOS has been experimentally probed several times \cite{Eisen, Smith, Field}, as has the tunneling DOS \cite{Turner}.  Countless other methods for calculating the bulk density of states for systems have been carried out, tailoring each model to meet specific material, dimensional or disorder constraints (see  \cite{Bahder, Bruno, Liboff, Gold, Anwar, Quang}). 

The investigations listed above, however, have rarely focused on the \emph{local} density of states. The bulk density of states measures the DOS averaged out across the sample; it returns a single value for every Fermi energy input. By contrast, the local density of states calculates the density of states independently at every given lattice point in the sample, returning thousands of values (each tied to a specific location) for every Fermi energy input. Researchers have oft preferred studying the bulk behavior, because, in the words of one group of authors, they were simply ``not interested in the details of the density of states." \cite{Kramer}. Yet, the local DOS is essential to understanding 1D systems \cite{Meyer}. 

There have been some local density of states calculations of note (see \cite{Jug}). Most relevantly, in 2003, Meyer \textit{et al} carried out the first measurement of the local density of states in an extended 1D system \cite{Meyer}. They measured the local DOS across a slice of the sample (a slice of a system will be more fully defined in subsequent chapters). Meyer then compared theoretical predictions for the local DOS based on single-particle calculations with measured values of the local DOS. To their surprise, for their measured values, they ``did not find significant deviations from the calculation." Theoretical studies of the local DOS are therefore seen to provide a potentially rich source of both accurate and fundamental information regarding 1D electron systems.

\begin{figure}
\centering
\includegraphics[width=0.9\textwidth, viewport= 40 300  800 710]{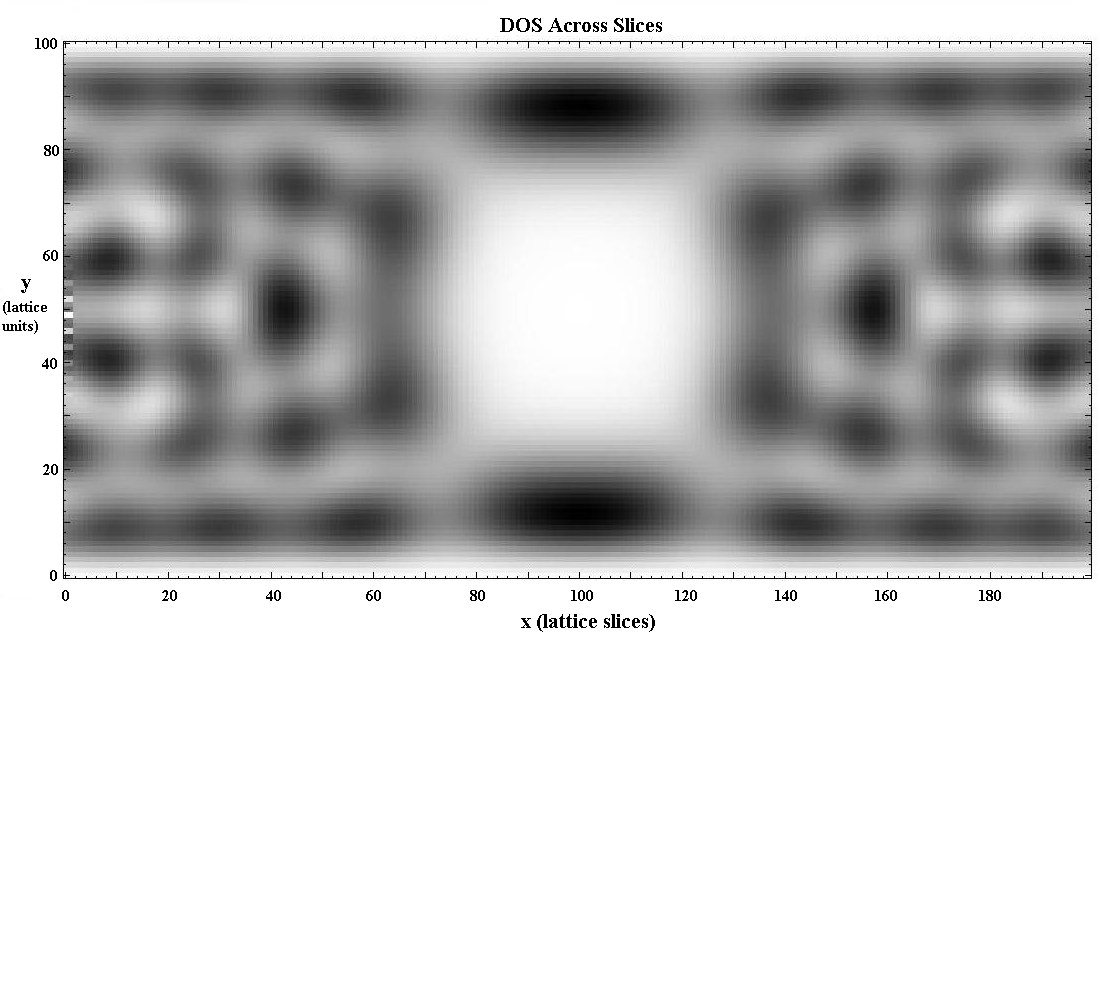}
\caption[Local Density of States Plot]{\textit {Two-Dimensional Local Density of States plot for a 1DEG under the influence of a central surface gate. This plot is calculated using the numerical Green's Function model presented in this thesis. The system has width 101 lattice points and length 200 lattice points and is at Fermi Energy 11 meV and field B =1 T. Black represents high density of states, white low. Depletion under the gate and the single subband that manages to pass along the edge are clearly visible.}}\label{fig:DOSSample}
\end{figure}

This thesis will investigate the behavior of the one-dimensional, local density of states using a numerical approach. The work contained herein will concern itself not merely with the local DOS across the 1D channel (as Meyer did), but simultaneously with the local DOS along the channel (in the direction of transport). Experiments on low-dimensional structures can only give conductance measurements, so one has to work backwards from these results in order to grasp the underlying physics. Yet, a numerical model---one that can compare conductance values and local density of states plots---provides a window into the \emph{structure} of the system: into scattering effects, transport, magnetic field response, and imaginary band structure. In short, the local DOS gives a highly detailed portrait of transport (see Figure \ref{fig:DOSSample}). With the aim of advancing knowledge about 1D systems, this thesis derives and presents numerical models for finding both the local DOS and the transmission coefficients. The results confirm theoretical predictions with a high degree of accuracy. The model can be deployed to aid numerous inquiries and applications.

This thesis contains four parts. Chapter Two provides an introduction to transport in 1DEG systems. In particular, the Landuaer-B\"{u}ttiker formalism is outlined and so is the unifying work of Baranger and Stone. The third chapter describes the numerical method to be implemented, a method based chiefly on a technique forged by MacKinnon using Green's functions \cite{MacDaddy}. In the fourth chapter, a detailed description of the local DOS and transmission programs that are the products of this thesis will be presented. It is shown that the results from these numerical models match expected theoretical values for changes in magnetic field and gate voltages with robust accuracy. Lastly, in Chapter Five, a selection of applications of the model is offered.

 \chapter{Quantum Transport Theory}
\section{Introduction}

To develop the numerical models presented later in this thesis, especially the transmission coefficients programs, a theoretical overview of quantum transport theory is required. This chapter begins with the basics of transmission in a perfect 1D channel, then advances to the two-probe Landauer conductance formula that can calculate transmission even in the presence of imperfections in the system. From this two-probe form, the multi-probe Landauer-B\"{u}ttiker formula is derived. The chapter then focusses on the work of Baranger and Stone who demonstrated the equivalence of the Landauer-B\"{u}ttiker scattering formalism and the exact eigenstate (and Green's functions) formalism.  Their work leads this thesis into a discussion of Green's functions (Chapter Three) which form the basis of the numerical model (Chapter Four). 

\section{Perfect 1D Transport}

Classically, electron transport in metals was described by the Drude Conductivity.  This equation,  in which $n_{s}$ is the sheet carrier density, $m^*$ is the electron's effective mass, $\tau_{e}$ is the mean free time, and $\mu_{e}$ is the electron's mobility, is given by: 

\begin{equation}
\sigma=\frac{n_se^2\tau_e}{m^*}=en_s\mu_e
\end{equation}

Though this equation works reasonably well in describing the transport for a 2DEG, it is not as effective for a 1DEG. The reason for this difference is that the dimensions of a 2DEG are typically much larger than the mean free path of the electron, meaning that one can use the average quantity $\tau_e$ with reasonable accuracy. Not so, however, for a 1DEG, where the width---and often the length, too---of the channel is typically less than the electron's mean free path. Since the electrons are usually confined electrostatically (e.g. by split gates), little to no scattering takes place along the width of the system. Longitudinal momentum is conserved. 

Since the Drude conductivity cannot accurately describe conductance in a 1DEG, a new formalism is required (see section \ref{LandSec}). Before deriving this general new method, however, it is necessary to develop a description of basic transport in a perfect 1D system by focussing on current flows in each direction. Assuming a perfect 1D channel, the current flowing in one subband, $i$, does not scatter into any other, but flows cleanly through the system at its subband energy. If one considers a system with a chemical potential greater on its right-hand contact than its left-hand contact by an energy $eV$, then the current flowing to the right in channel $i$ is given by: 

\begin{equation}
dI_{i}^{+}=-ev_{\epsilon}^{i}f(\epsilon+eV)\frac{dn_{i}}{d\epsilon}d\epsilon
\end{equation}
within an energy range $d\epsilon$. $v_{\epsilon}^{i}$ is the group velocity at energy $\epsilon$, $f(\epsilon+eV)$ is the Fermi-Dirac function which gives the probability of an electron being in a given state, and $\frac{dn_{i}}{d\epsilon}$ is the density of states, again at energy level $\epsilon$. One of the crucial features of this expression is that since $v_{\epsilon}^{i}\sim\frac{dE_{k_{x}}}{dk_x}$ and $\frac{dn_{i}}{d\epsilon}\sim\frac{dk_x}{dE_{k_{x}}}$, the group velocity and density of states terms drop out. The system's dependence on energy, subband number, and momentum vanish from the expression for current, leaving only the Fermi-Dirac distribution function and constants:

\begin{equation}
dI_{i}^{+}=-\frac{e}{h}f(\epsilon+eV)d\epsilon
\end{equation}

Meanwhile, by an identical process, current flowing from the right-hand contact to the left-hand contact is simply:

\begin{equation}
dI_{i}^{-}=-\frac{e}{h}f(\epsilon)d\epsilon
\end{equation}

The net current flowing through energy subband $i$ is the difference between the right-flowing and left-flowing currents integrated across the full energy range: 

\begin{equation}
I_i=\frac{e}{h}\int_{-\infty}^{\infty}(f(\epsilon)-f(\epsilon+eV))d\epsilon
\end{equation}

Since the Fermi-Dirac function is given by:

\begin{equation}
f(\epsilon)=\frac{1}{e^{\frac{\epsilon-\mu}{kT}}+1} 
\end{equation}
where $\mu$ is the chemical potential, $k$ is Boltzmann's constant, and $T$ is temperature, the function becomes a right-angled step function when the temperature is taken to zero, yielding the final transmission result: 

\begin{equation}
I_i=\frac{e^2}{h}V
\end{equation}

For $N$ occupied subbands in the 1D system, the conductance is:

\begin{equation}
G=\frac{dI}{dV}=N\frac{e^2}{h}
\end{equation}

The above result is a fundamental theoretical insight into electronic behavior. It does not, however, provide detailed information about transport beyond a perfect 1D system. A lucid and simple formalism adaptable to studying wide-ranging quantum electronic systems including 1DEGs is the Landauer-B\"{u}ttiker method presented below.

\section{The Landauer Method}\label{LandSec}

The novel insight that earned Landauer his eponymous formula was the possibility of recasting a conductance problem as a scattering one \cite{LandOne, LandTwo}. Instead of concentrating on the effect of applied electric fields, Landauer zeroed in on the transmission and reflection coefficients across channels in a quantum system \cite{Szafer}. If one considers a pair of 1D electron systems, attached on either side to perfect Ohmic contacts and with an arbitrary potential region lying in between (see Figure \ref{fig:twodchannel}) the two-probe Landauer formula can be derived straightforwardly. As was the case for the transport system analyzed above, a chemical potential on the right-side contact is greater by energy $eV$ than the left-side contact. The current amplitude crossing the system to the right is positive and denoted by $a^+$ and $b^+$, while the current amplitude moving to the left is negative and given by $b^-$ and $a^-$. The total current amplitude that flows into the right contact, $b^+$, and left contact, $a^-$, can be written in matrix form:

\begin{figure}
\centering
\includegraphics[width=0.9\textwidth, viewport= 60 290 380 380]{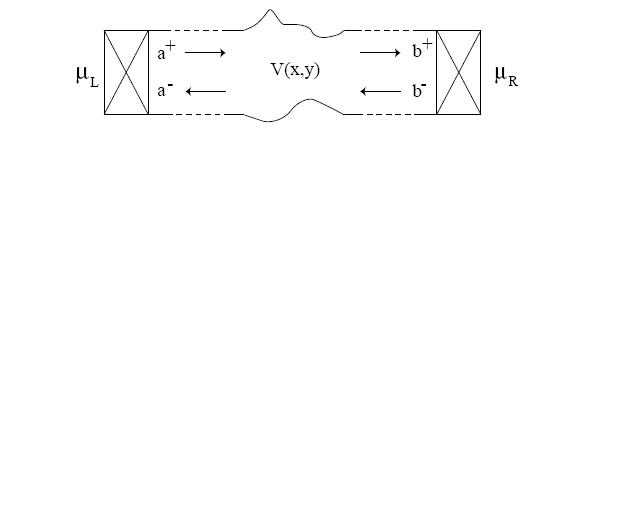}
\caption[Quantum Channel with Leads]{\textit{Quantum channel with leads attached. Positive current moves right, negative current left, under the influence of the left lead's chemical potential, $\mu_L$, the right lead's chemical potential, $\mu_R$, and passing through a quantum transport region with arbitrary effective potential, V(x,y).}}\label{fig:twodchannel}
\end{figure}

\begin{equation}
\left(\begin{array}{cc}
b^+\\
a^-\end{array}\right) =
\left(\begin{array}{cc}
t^+ & r^+\\
r^- & t^- \end{array}\right)
\left(\begin{array}{cc}
a^+ \\
b^- \end{array}\right) 
\end{equation}
where $t^+, t^-, r^+,$ and $r^-$ are the transmission and reflection matrices.

One can see that $b^+$ is simply the sum of transmitted portion of $a^+$ and the reflected portion of $b^-$, while $a^-$ is simply the sum of the reflected portion of $a^+$ and the transmitted portion of $b^-$. Each of these amplitudes, $a^+, a^-, b^+, b^-$, are themselves vectors, their entries being, for example, $a^+_1, a^+_2, a^+_3,$ which are the amplitudes of a single subband $i=$1, 2, and 3. Current transmitted to the right, in a narrow energy range $d\epsilon$, is then determined by: 

\begin{equation}
dI^+=\langle a^{+\dag} t^{+\dag} t^+a^+\rangle
\end{equation}
with the above expression averaged over time. This expression represents the current that has crossed the system. The time average of $\langle\vert a^+_i\vert^2\rangle$ is simply the familiar $-\frac{e}{h}f(\epsilon+eV)d\epsilon$, and therefore the equation for $dI^+$ is given by: 

\begin{equation}
dI^+=-\frac{e}{h}Tr[t^{+\dag} t^+]f(\epsilon+eV)d\epsilon
\end{equation}

The trace of $t^{+\dag} t^+$ is calculated in order to sum the transmission coefficients for each subband. Qualitatively, this sum makes sense, since the above equation is essentially calculating the portion of the current that is transmitted. By similar argument to above, for the right side of the system, one obtains: 

\begin{equation}
dI^-=-\frac{e}{h}Tr[t^{+\dag} t^+]f(\epsilon)d\epsilon
\end{equation}

Using $G=dI/dV$, integrating for all energy levels, and once more taking the zero-temperature limit, one obtains the Landauer formula: 

\begin{equation}\label{LanCorrect}
G=\frac{e^2}{h}Tr[t^{+\dag} t^+]=\frac{e^2}{h}\sum\vert t_{ij}\vert^2
\end{equation}
where $t_{ij}$ is the conductance coefficient for charge moving from subband $i$ on the left-side portion of the quantum system into subband $j$ on the right-side portion of the system. This derivation assumes an arbitrary effective potential and therefore applies to any 1D system connected at both ends to ohmic contacts. In the ideal situation, these ohmic contacts behave like electron-emitting blackbody radiators. 

\section{The Challenge of the Multi-Probe Formalism}

Landauer's initial 1957 paper determined the conductance formula to be:

\begin{equation} \label{LanWrong}
G=\frac{e^2}{h}\frac{T}{R}
\end{equation}
For decades this formula stood untouched \cite{Szafer}. It was only when Anderson \cite{Anderson}, in 1980, and others \cite{Economou, FisherLee, Sivan, Langreth, Eng} attempted to derive a \emph{multi-channel} version of Landauer's equation that a conductance equation with $G$ proportional only to $T$  (\ref{LanCorrect}) and not $G\sim T/R$ was derived \cite{Economou, FisherLee}.

Part of the concern with formulas like (\ref{LanCorrect}) had been that it gave a finite resistance even for a perfect conductor. This finite resistance proved to be the result of the contact resistance: a resistance resulting from electrons with momentum distributions that did not match the quantized levels allowed in the channel trying to enter the 1D system \cite{ImryBook, LandZPB}. The attempt to develop a multi-channel formalism also grappled with the issue of whether the conductance measured at fixed current was a measure of the chemical potential at a set of charge reservoirs (source and sink far away from the system) or at an ``effective chemical potential" inside the sample \cite{Szafer}. Ultimately, it was agreed that despite transmission being measured \emph{within} the channel, it was acceptable to use the chemical potentials of the reservoirs \emph{outside} the channel. This idea was demonstrated by B\"uttiker and is derived below \cite{ButtPRL}. 

\section{The Landauer-B\"uttiker Formalism}

B\"uttiker put forward the successful multi-probe version of the Landauer theory by treating current and voltage terminals in a four-point probe set-up equally \cite{ButtPRL, ButtIBM}. Swapping current and voltage probes, he demonstrated that such systems obeyed Onsager's relations \cite{OnsagerOne, OnsagerTwo} (see \cite{ButtPRL} for derivation of Onsager's relations).

B\"uttiker considered four reservoirs, each at different chemical potential $\mu_i$ with a fifth chemical potential, $\mu_0$ less than or equal to the lowest of all four $\mu_i$. Since states with energy below $\mu_0$ are filled,they cannot contribute any net current to the leads. Therefore, the only relevant energy range is $\Delta\mu_i=\mu_i-\mu_0$ above $\mu_0$. The current injected by reservoir $i$ into the system is, as was derived above: 

\begin{equation}
I_i=\frac{e}{h}eV=\frac{e}{h}\Delta \mu_i
\end{equation}
Part of the current is also reflected back into the reservoir. The magnitude of the proportion of current reflected is determined by a reflection coefficient, $R_{ii}$,  and results in: 

\begin{equation}
I_{i(Reflected)}=-\frac{e}{h}R_{ii}\Delta\mu_i
\end{equation}

Lastly, current is also flowing into the reservoir from other reservoirs. For each reservoir besides $i$ ($i+1, i+2,$ etc.), there will be a transmission current with magnitude given by a transmission coefficient, e.g. $T_{12}$ for current injected into the system from reservoir 2 that ends up in reservoir 1. Thus the impact on reservoir 1 from three other leads is given by:

\begin{equation}
I_{i(Transmitted)}=-\frac{e}{h}\left(T_{12}\Delta\mu_2+T_{13}\Delta\mu_3+ T_{14}\Delta\mu_4\right) 
\end{equation}
In general, the net current flowing out of a given lead is therefore determined by the multi-probe Landauer-B\"uttiker equation:

\begin{equation}\label{ButtStuff}
I_i=-\frac{e}{h}\left((1-R_{ii})\mu_i-\sum_{i\neq j}T_{ij}\mu_j\right) 
\end{equation}

The $\mu_0$ terms cancel out because the coefficients  sum to zero. This can be seen clearly when (\ref{ButtStuff}) is written out as an $N\times N$ matrix, where N is the number of reservoirs. For a $3\times 3$ system this would look like: 

\begin{equation}
\left(\begin{array}{ccc}
I_1\\
I_2 \\
I_3 \end{array}\right)
= \left(\begin{array}{ccc}
1-R_{11} & -T_{12} & -T_{13} \\
-T_{21} & 1-R_{22} & -T_{23} \\
-T_{31} & -T_{32} & 1-R_{33}\end{array}\right) 
\left(\left(\begin{array}{ccc}
\mu_1 \\
\mu_2 \\
\mu_3 \end{array}\right) \\
-\left(\begin{array}{ccc}
\mu_0 \\
\mu_0 \\
\mu_0
\end{array}\right)\right)
\end{equation}
Every column and every row of the $3\times 3$ matrix sums to zero, $1-R_{ii} -\sum_j T_{ij}=0$, because physically speaking if the energy in every lead were the same, no current would flow. Thus, when any row is multiplied by a vector with identical entries---the $\mu_0$ vector---the result will be zero. This method of eliminating $\mu_0$ obviated the need for a ``local chemical potential," allowing potentials to be calculated at the reservoirs alone. More generally, for $N$ occupied subbands in lead $i$, the equation takes the form: 

\begin{equation}
I_i=-\frac{e}{h}\left((N_i-R_{ii})\mu_i-\sum_{i\neq j}T_{ij}\mu_j\right)
\end{equation}

This is the multi-probe formalism for an arbitrary quantum system with an arbitrary number of leads. Knowing only the reservoir chemical potentials, the transmission and reflection coefficients, and the number of conducting subbands, one can calculate the conductance in any quantum system. This result is essential to the numerical model used to find the transmission and reflection coefficients implemented in Chapter Four. 

\section{The Baranger and Stone Unification}

\begin{figure}
\centering
\includegraphics[width=0.9\textwidth, viewport= 20 120 380 360]{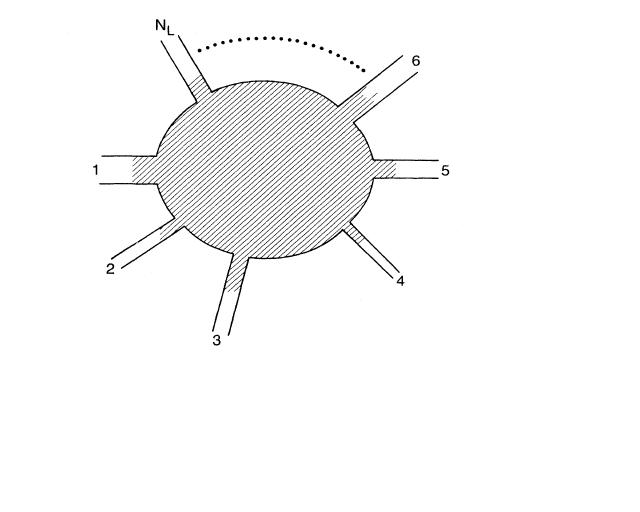}
\caption[Quantum System with N Leads]{\textit{Quantum system with arbitrary number of leads. The shaded region is the arbitrary-geometry area of quantum interaction. Image from \cite{Baranger}.}}\label{fig:BandSDiagram}
\end{figure}

The final key piece of transport theory is the work of Baranger and Stone \cite{Baranger}. They showed the equivalence of the two chief approaches to linear response theory ($I \sim V$), which are also the two approaches required for this thesis's numerical models: the Landauer-B\"uttiker scattering formalism (transmission program) and the exact eigenstate, Kubo-Greenwood, Green's function formalism (local DOS and transmission programs). Baranger and Stone made no assumptions except current-conservation, time-reversal symmetry, and the non-interacting electron model. They began by finding the exact eigenstate form of the conductance coefficients, $g_{mn}$ that solve the linear equation $I_m=\sum_n g_{mn}V_n$, where leads $m$ and $n$ inject current into a quantum system with an arbitrary number of leads, $N_L$ (see Figure \ref{fig:BandSDiagram}). Then, they related $g_{mn}$ to the Landauer-B\"uttiker transmission coefficients $T_{mn}$.

They, following B\"uttiker, shunned the idea of ``an effective chemical potential" lying somewhere inside the conductor. This ``effective" or ``local chemical potential," which plagued previous research had been developed in order to overcome the issue of particles scattering into voltage probes. Yet, conductors are necessarily out of equilibrium if current is flowing, and hence theories about equilibration of different channels at the Fermi energy tended to be arbitrary. Baranger and Stone concerned themselves solely with current injected into the system. All that was required were the appropriate boundary conditions: a reservoir that was in equilibrium at fixed potential $\mu$, that was large enough so its potential would be unchanged by an additional particle, that inelastically scattered (phase-randomized) any particle entering the reservoir before returning it to the system, and that had a boundary with the sample that caused no additional resistance. 

Their derivation of $g_{mn}$ was carried out not by using conductivity, a spatially average quantity, but with the conductivity response function, $\sigma(x, x')$. Since small spatial-fluctuations can greatly impact mesoscopic systems, it is important that $\sigma(x, x')$ is a \emph{spatially-varying} quantity describing the current density. It is also a function of states both at and below the Fermi surface;  consequently, Baranger and Stone proved that the transport current is only a Fermi surface entity. Moreover, also unlike previous research \cite{Szafer, FisherLee} their derivation was the first to apply in a magnetic field of arbitrary strength. 

\begin{figure}
\centering
\includegraphics[width=0.9\textwidth, viewport= 20 170 380 380]{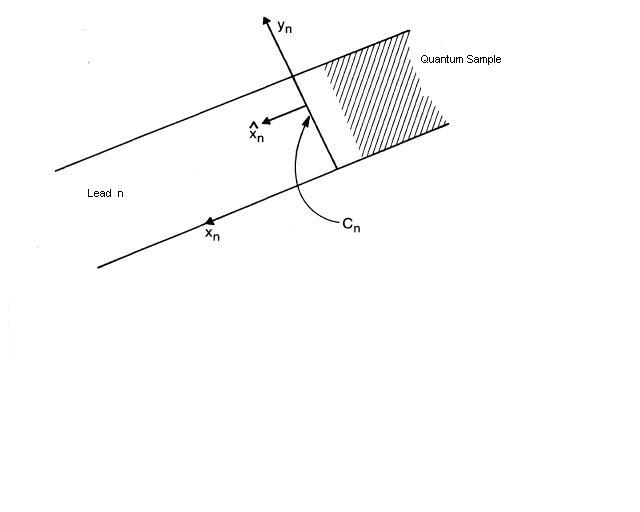}
\caption[Sample-Lead Boundary]{\textit{Transition from quantum sample into lead $n$. $x_n$ points in the direction of the lead $n$; $y_n$ is transverse to it. $C_n$ is the cross section of lead $n$, perpendicular to $\hat x_n$. Image from \cite{Baranger}.}}\label{fig:BandSDiagramLead}
\end{figure}

The conductance coefficients $g_{mn}$ are identified as (see Figure \ref{fig:BandSDiagramLead}):

\begin{equation}\label{leadStuff}
g_{mn}=-\int_{C_m}dy_m\int_{C_n}dy_n' \mathbf{\hat x}_m\cdot\sigma(x,x')\cdot\mathbf{\hat x}_n 
\end{equation}
where $\mathbf{\hat x}_n$ is the unit vector parallel to lead $n$, $y_n$ points perpendicular to the lead, and $C_n$ is the cross section of the lead. As such, equation (\ref{leadStuff}) gives a physically intuitive expression for $g_{mn}$: it is the flux of the conductivity-response function from lead $n$ into lead $m$ that passes through the cross sections of each lead, $\int_{C_m}dy_m$ and $\int_{C_n}dy_n'$. The full calculation of $\sigma(x,x')$ is a laborious task whose expressions can be found as exact eigenstates or Green's functions in equations (40) and (75), respectively, in \cite{Baranger}. 

The conductance coefficients, $g_{mn}$, are then manipulated so they are related to  $t_{mn}$, the subband transmission coefficients, by:

\begin{equation}
g_{mn}=\frac{e^2}{h}\int d\epsilon\left(-\frac{df}{d\epsilon}\right)\sum_{a,c} \vert t_{mn,ca}\vert^2, \hspace{5mm} m\neq n 
\end{equation}
where $c$ is a subband in lead $m$, $a$ is a subband in lead $n$, and $-df/d\epsilon$ is the derivative of the Fermi function. Taking the zero-temperature limit, one obtains:

\begin{equation} 
g_{mn}=\frac{e^2}{h} T_{mn}, \hspace{8mm} m\neq n
\end{equation}
where $T_{mn}$ is the trace of the $t^{+\dag}t^+$ matrix.

In this way, the exact eigenstate solution of linear-response theory stemming from $\sigma(x,x')$ and the scattering formalism derived from transmission and reflection coefficients are seen to be equivalent. Quantum transport calculations performed with exact eigenstates are identical to calculations performed with transmission and reflection coefficients. Each form has its own advantages. Green's Functions, which are derived from the exact-eigenstate formalism, are powerful numerical tools, while transmission coefficients give a physically intuitive description of transport. The method of using Green's functions for numerical simulations---the backbone of the programs of this thesis---is the subject of the next chapter.

\chapter{Numerical Green's Functions}

\section{Introduction}
Green's functions are used to solve inhomogeneous differential equations and provide an effective method for analyzing the local density of states, conductance, and other transport-related properties of semiconductor systems. This chapter explores how they can be used to create a numerical model. It begins with an analysis of the discrete lattice (as opposed to continuous wave functions) and then considers the application of appropriate effective potentials to the system. This chapter then moves into an analysis of Green's functions: their definition and their utility, how they are used to solve a Hamiltonian system, how they are used iteratively to calculate transport properties, and lastly the appropriate boundary conditions. The iterative process, it should be emphasized, is crucial, as it allows one to calculate the energetics of the entire system (e.g. 100,000 lattice points) by repeatedly multiplying only vertical slices of the system (each with only 100 lattice points); this permits the multiplication of matrices of the order of $100 \times 100$ rather than having to perform an inversion of a $10^5 
\times 10^5$ matrix---a massive numerical task.

\section{Creating A Discrete System}
\subsection{Discretizing the Schr\"{o}dinger Equation}

In order to create a matrix representation of a quantum system, the relevant Schr\"{o}dinger equation must be discretized. To allow for maximum generality, a magnetic field is applied, perpendicular to a two-dimensional system, with magnetic vector potential, $A$. The Landau gauge, $A=\langle-By, 0,0\rangle$, is used. Employing the Peierls substitution of inserting $eA$ into the Hamiltonian \cite{Luttinger, Peierls}, the Schr\"{o}dinger equation for the two-dimensional system is determined by: 

\begin{equation}
\frac{1}{2m^*} \left( -i \hbar \frac{\partial}{\partial x}+eBy \right)^2 \psi
-\frac{\hbar^2}{2m^*}\frac{\partial^2}{\partial y^2}\psi + V\psi = E\psi
\end{equation}

A lattice constant, $a$, is introduced. This transforms the equation into: 

\begin{equation}\label{Hamil2}
-\left( a \frac{\partial}{\partial x}+ \frac{ieBya}{\hbar} \right)^2 \psi
-\left(a\frac{\partial}{\partial y}\right)^2\psi + \frac{2m^*a^2}{\hbar^2}V\psi = \frac{2m^*a^2}{\hbar^2}E\psi 
\end{equation}

The squared terms in (\ref{Hamil2}) are replaced by a second order Taylor polynomial for $e^x+e^{-x}$ using $x^2\approx e^x+e^{-x}-2$: 

\begin{equation}
4\psi-\left(e^{a\frac{\partial}{\partial x}}e^{\frac{i \gamma y}{a}} +e^{-a\frac{\partial}{\partial x}}e^{-\frac{i \gamma y}{a}}\right)\psi
-\left(e^{a\frac{\partial}{\partial y}}+e^{-a\frac{\partial}{\partial y}}\right)\psi+\frac{2m^*a^2}{\hbar^2}V\psi=\frac{2m^*a^2}{\hbar^2}E\psi 
\end{equation}
where $\gamma=\frac{eBa^2}{\hbar}$.

\begin{figure}
\centering
\includegraphics[width=0.9\textwidth, viewport= 20 260 460 600]{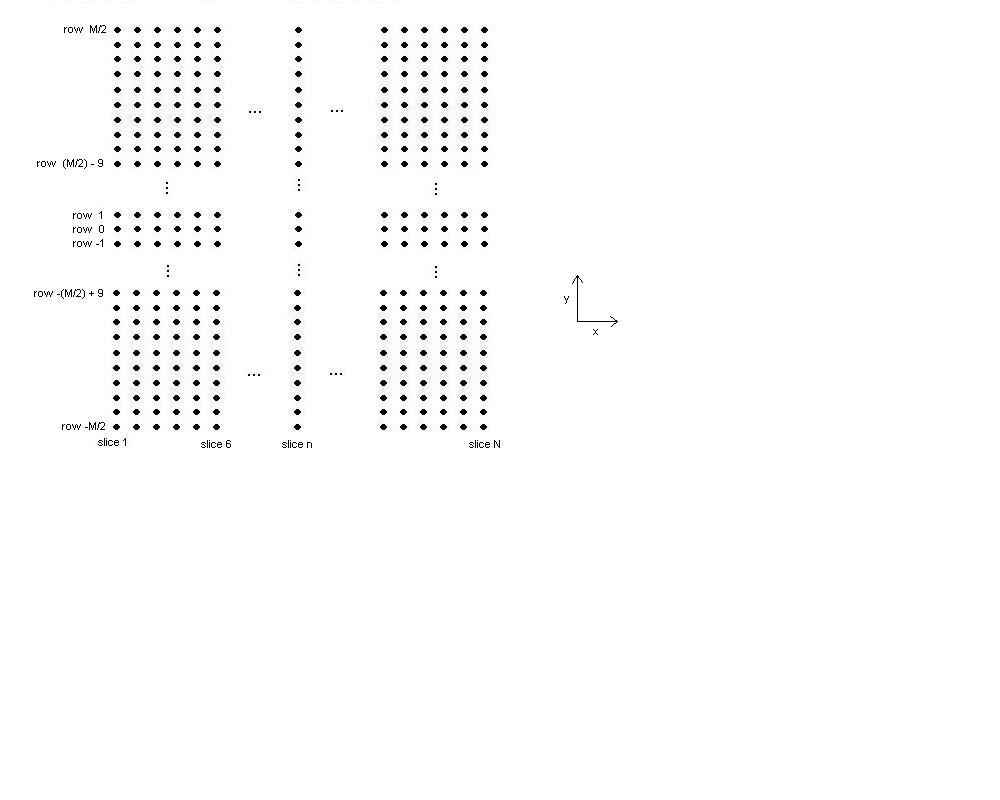}
\caption[Computational Lattice]{\textit{Lattice used for computation. Each dot represents a lattice point. The numerical model is carried out by multiplying matrices representing vertical lattice slices (e.g. slice $n$). Positive current is taken to be flowing in the positive x-direction.}}\label{fig:LatticeMap}
\end{figure}

Now the system is made discrete by deploying a Taylor expansion once again. Using the fact that, to second order:

\begin{equation}
e^{a\frac{\partial}{\partial x}}(\psi(x))\approx\psi(x)+a\frac{\partial\psi}{\partial x}+\frac{a^2}{2}\frac{\partial^2\psi}{\partial x^2}
\end{equation}
and that
\begin{equation}
\psi(x+a)\approx\psi(x)+a\frac{\partial\psi}{\partial x}+\frac{a^2}{2}\frac{\partial^2\psi}{\partial x^2}
\end{equation}
one sees that $e^{a\frac{\partial}{\partial x}}(\psi(x))\approx\psi(x+a)$. The corresponding result $e^{-a\frac{\partial}{\partial x}}(\psi(x))\approx\psi(x-a)$ also holds. Replacing $x$ and $y$ with lattice points $n$ and $m$ related by $x=na$ and $y=ma$ (see Figure \ref{fig:LatticeMap}), the discretized Hamiltonian is obtained: 
\begin{equation}\label{eq:discretized}
e^{i\gamma m}\psi_{n+1,m} + e^{-i \gamma m}\psi_{n-1,m} + \psi_{n,m+1} + \psi_{n,m-1} +\nu\psi_{n,m}=\epsilon\psi_{n,m}
\end{equation}
where $\nu=-\frac{2ma^2}{\hbar^2}V$, $\epsilon= 4-\frac{2ma^2}{\hbar^2}E$, and  $\psi_{n+1,m}$ represents  the wavefunction one lattice point right of the wavefunction at point $(n,m)$. Using standard error approximation methods for a Taylor series for $e^{a\frac{\partial}{\partial x}}$, the error will be no greater than $R_n=\frac{a^3}{6}e^\xi$ where $\xi$ is less than $a$, and for $e^{-a\frac{\partial}{\partial x}}$ the error will be no greater than $R_n=\frac{a^3}{6}$. Typically, $a$ is of length 5 nm, making these errors very small indeed. As will be shown in Chapter Four, the numerical model's results and the results expected by theory match extremely closely.

This discrete result is known as the tight-binding approximation. If one inserted $\psi_{n,m}=e^{ik_xan}e^{ik_y am}$ as a solution and set $\nu=0$, one would end up with cosine bands as the dispersion curves: $E=2\cos(k_xa)+2\cos(k_y a)$. Thus the solutions are tightly bound in the curvature of the cosine bands. Having found a discrete form of the Hamiltonian (\ref{eq:discretized}), the system can now be translated into matrix form and the method of numerical Green's functions described. First, however, the value of the effective potential,  $\nu$, is calculated. 

\subsection{Calculation of the Effective Potential}
To calculate the effective potential in a quantum system, the most thorough mechanism would be to calculate the potential self-consistently. This, however, is not the method used here. Self-consistent potentials require the simultaneous solution of the Schr\"{o}dinger and Possion charge distribution equations, a numerically intensive process. As such, though self-consistent potentials produce useful results (see, for example \cite{Laux}), they do not well serve a model designed for rapid adaptation to a large range of geometries and surface gates in quantum systems. 

Instead, a powerful and easily malleable tool for calculating the potential can be found in the work of Davies \textit{et al} \cite{DaviesL}. Their model is specifically tailored to measuring the effects of gates placed on the surface of a 2DEG. They do not factor in the contribution of the fields generated by the electrons themselves, but their results are nonetheless very practical and accurate for numerous reasons detailed in \cite{DaviesL}.

Their model derives from the solution to Laplace's equation, $\nabla^2\phi=0$. The first boundary condition is that $\phi (\textbf{r},0)$, the potential, is equal to $V_g$, the applied gate voltage, and the depth $z=0$ is the surface where gates are put down. The second condition is that $\partial \phi / \partial z=0$ in the limit $z\rightarrow \infty$. Then the two-dimensional Fourier transform is applied to $\phi(\textbf{r}, 0) $ turning it into $\tilde {\phi} (\textbf{q}, 0)$. Given that $z$ must decay exponentially in order to satisfy $\partial\phi/\partial z=0$ as $z\rightarrow \infty$, the general expression for the transform is given by: $\tilde{\phi}(\textbf{q},z)=\tilde{\phi}(\textbf{q},0)e^{-\mid qz \mid}$. This multiplication of the Fourier Transform, of course, is the same as convolution in real space. As a result, taking the inverse Fourier Transform yields:

\begin{equation}
\phi(\textbf{r}, z)=\int \frac{\vert z \vert}{2 \pi (z^2 + \vert \textbf{r} - \textbf{r\texttt{\char"0D}} \vert^2)^{\frac{3}{2}}} \phi (\textbf{r},0)d\textbf{r\texttt{\char"0D}}
\end{equation}

This general equation can then be manipulated for a whole host of results. The geometries of the gates simply need to be given in polar coordinates $\textbf{r}=(r,\theta)$. In general, the results are $\arctan(x,y)$ functions, resulting from the integral across the surface of the 2DEG.

The most common gate deployed in this thesis's calculations was a finite rectangular gate. Its effective potential is given by:

\begin{eqnarray}
\frac{\phi (\textbf{r}, d)}{V_g} & = &  g(x-L, y-B) +  g(x-L, T-y) \\
& + & g(R-x, y-B) +g(R-x, T-y) 
\end {eqnarray}
where $g(i,j)=\frac{1}{2\pi} \arctan (\frac{ij}{dR})$ and $R=\sqrt{i^2+j^2+d^2}$. The depth of the 2DEG below the surface is $d$, and $L, R, B, \mbox{ and } T$ are the values of the left, right, bottom, and top edges of the rectangular gate. Naturally, one could create an arbitrary number of gates and simply sum their effective potentials by the power of the superposition principle.

While the Davies \textit{et al} formulation allows for the calculation of properties for gate designs of all varieties and geometries, certain calculations are best carried out with a confining potential free of surface gates. In such a scenario, the confining potential of an infinite square well or a simple harmonic oscillator can be used with effective results (the potentials being set up transverse to the current). Their numerical implementation is discussed briefly in the next chapter. The saddle point potential---$\phi(x,y)=\phi_0-\frac{1}{2}m\omega_x^2x^2+\frac{1}{2}m\omega_y^2y^2$---is another very effective model for the potential arising from a split-gate \cite{Saddle}.

The groundwork for the numerical method has been laid by developing a discretized quantum lattice and an effective potential. The Green's function numerical technique is now presented.

\section{The Green's Function Numerical Method}
\subsection{A Green's Function Primer}
\subsubsection{The Definition}

Green's functions are implemented to solve inhomogeneous differential equations. Consider a partial differential equation of the form: 

\begin{equation}
\mathcal{L}y(r)=\mathcal{F}(r)
\end{equation}
where $\mathcal{L}$ is a linear operator on $y(r)$ and $\mathcal{F}$ is the inhomogeneity. The solution, $y(r)$, is written in terms of the Green's function, $G(r, r')$, and its product with the inhomogeneity: 

\begin{equation}
y(r)=\mathcal{L}^{-1}\mathcal{F}(r)=\int G(r, r')\mathcal{F}(r')dr'
\end{equation}

Hence, $\mathcal{F}(r)=\int \mathcal{L} G(r, r')\mathcal{F}(r')dr'$, and as a consequence of the definition of the dirac-delta function: 

\begin{equation}
\mathcal{L}G(r, r')=\delta(r-r')
\end{equation}

The Green's function can be thought of as the inverse of the linear operator, as is especially clear when the differential equation is cast in matrix form, $\mathbf{\mathcal{L}G}=\mathbf{I}$, where $\mathbf{I}$ is the identity matrix.

\subsubsection{The Eigenvalue Equation and $s$}
Equation (\ref{eq:discretized}) can be recast as the energy eigenvalue equation $H\psi=\epsilon\psi$. This equation can be rewritten $(\epsilon-H)\psi=0$, and in this form, $\epsilon-H$ plays the role of $\mathcal{L}$. The Green's function is therefore given by:

\begin{equation}\label{Green1}
(\epsilon-H)G(r,r')=\delta(r-r')
\end{equation}

Normally, however, a complex energy $z=\epsilon+is$ is defined and used to replace $\epsilon$. $s$ is made infinitesimally small, in order to avoid having an impact on the numerical result \cite{MacDaddy}. The reason for the inclusion of $s$ can be understood when \ref{Green1} is rearranged for $G(r,r')$:

\begin{eqnarray}
G(r, r', z) & = & \frac{1}{z-H}\delta(r-r') \\
 & = &  \sum_n \frac{\psi_n(r) \psi_n^*(r')}{z-H} \\
 & =  & \sum_n \frac{\psi_n(r) \psi_n^*(r')}{z-\epsilon_n} \label{En}
\end{eqnarray}
Where the substitution of $\epsilon_n$ for $H$ in (\ref{En}) is made because $H \vert\phi_n \rangle=\epsilon_n \vert\phi_n \rangle$. As can be seen, since $H$ is Hermitian and therefore $\epsilon_n$ is real, $G(r,r')$ is analytic everywhere except at the eigenvalues of $H$. $G(r,r')$ has poles at these discrete eigenvalues, $\epsilon_n$. Consequently, to avoid the problem of having to calculate residues wherever $z=\epsilon_n$, $s$ is added to $\epsilon_n$. In calculations performed in Chapter Four, $s$ was set to $10^{-18}$: small enough to have no impact on the results, but large enough to avoid computational error. 

\subsubsection{The Dyson Equation}
One of the most important properties of Green's functions is their simple reformulation when a perturbation is added. Consider a system with solution $G_0=\frac{1}{z-H_0}=(z-H_0)^{-1}$. A perturbation with Hamiltonian $H_1$ is added. The total Hamiltonian is now given by $H=H_0+H_1$ and the total Green's function is: 

\begin{equation}
G=(z-H_0-H_1)^{-1}=(G_0^{-1}-H_1)^{-1}
\end{equation}

Multiplying both sides by the inverse of the right hand side and then by $G_0$ one arrives at $G-G_0H_1G=G_0$, or alternatively:

\begin{equation} 
G=G_0+G_0H_1G
\end{equation}
This relation is the well-known Dyson equation. Since adding a new lattice slice to the quantum system is adding a new perturbation, $H_1$, the Dyson equation plays a fundamental role in the iterative method described in section \ref{sec:iterative}.

\subsection{The Green's Function and the Hamiltonian}
The Hamiltonian of the system must account for every point in the $N \times M$ lattice (Figure \ref{fig:LatticeMap}). The total Green's function matrix for the whole system must therefore be of the same dimensions. Fortunately, calculating these enormous matrices is not required. Rather, if the system is divided up into $N$ vertical slices, each slice having $M$ lattice points, then the relevant matrices are the matrices of each slice, of dimension $M \times M$ instead of $NM \times NM$. Inverting an $NM \times NM$ matrix to solve for every entry of $\mathbf{G}$ would be an enormous numerical calculation for even a modest system of $N=500$ and $M=100$. Instead, the matrices for the Hamiltonian $\mathbf{H}$, the energy $\mathbf{Z}$, and the Green's functions, $\mathbf{G}$ are all made to correspond to only a single vertical slice. The matrix relationship relating all the lattice points on slice $i$ to all the lattice points on slice $j$ (see Fig. \ref{fig:LatticeMap}) is given by: 

\begin{equation}\label{NN}
\left[\mathbf{Z}-\mathbf{H}_{i,i}\right]\mathbf{G}_{i,j}-\mathbf{H}_{i,i+1} \mathbf{G}_{i+1,j}-\mathbf{H}_{i,i-1}\mathbf{G}_{i,j-1}=\mathbf{I}\delta_{i,j}
\end{equation}

Given the nearest neighbor approximation form of \ref{eq:discretized}, in which only the effects of neighboring lattice points impact the Schr\"{o}dinger equation for that lattice point, \ref{NN} is appropriate here. As a result, only three terms are present above: $z-H$ on the slice, $H$ one slice to the right, and $H$ one slice to the left. For clarity, it helps to write out $\left[\mathbf{Z}-\mathbf{H}_{i,i}\right]$. On slice $n$, with effective potential at point $m$ given by $\nu_{n,m}$, $\left[\mathbf{Z}-\mathbf{H}_{i,i}\right]$ is: 

\begin{equation}\label{ZminusH}
\left(\begin{array}{cccccc}
z-\nu_{n,-\frac{M}{2}} & -1 & 0& \ldots &  \ldots &\ldots \\
-1 & z-\nu_{n,-\frac{M}{2}+1} & -1 \\
0 &  &\ddots \\
\vdots &  &   -1 &  z-\nu_{n,0} & -1 &\\ 
\vdots &  &  &  &   \ddots & \\
\vdots &  &  &  & -1 & z-\nu_{n,\frac{M}{2}}
\end{array}\right)
\end{equation}

The above can be thought of as the matrix representation form of (\ref{eq:discretized}), where each row of the matrix corresponds to the Schr\"{o}dinger equation for a particular lattice point located at position $(n,m)$ in the lattice. The $e^{i\gamma m}\psi_{n+1,m}$ and $e^{-i \gamma m}\psi_{n-1,m}$ terms are reserved for the $\mathbf{H}_{i,i+1}$ and $\mathbf{H}_{i,i-1}$ matrices, respectively. There are a few features worth noting in (\ref{ZminusH}). First, $\epsilon$ has been replaced by complex $z$. Second, the off-diagonal -1 terms represent, reading across a row, the $\psi_{n,m-1}$ and $ \psi_{n,m+1}$ terms of Hamiltonian. Thirdly, the effective potential is centered around zero, from lattice position $-\frac{M}{2}$ to $\frac{M}{2}$, instead of from 0 to $M$. 

The reason for centering each slice's label around zero is to make the matrices symmetric as is evident in the forms of the other two Hamiltonian matrices in (\ref{NN}). The Hamiltonian linking one slice to its neighbor on the right is given by:

\begin{equation}\label{V}
\mathbf{H}_{i,i+1}=\mathbf{V}=\left(\begin{array}{cccccc}
e^{-i \gamma \frac{M}{2}} & 0 & 0& \ldots &  \ldots & 0  \\
0 & e^{i \gamma (-\frac{M}{2}+1)} & 0 \\ 
0 & 0 &\ddots \\
\vdots &  &  & 1\\
\vdots &  &  &  & \ddots\\
0 &  &  &  &  & e^{i \gamma\frac{M}{2}} 
\end{array}\right)
\end{equation}
where, as before, $\gamma=\frac{eBa^2}{\hbar}$. The third contribution to (\ref{NN}) is defined by $\mathbf{H}_{i,i-1}=\mathbf{V^\dag}$. In both $\mathbf{V}$ and $\mathbf{V^\dag}$, the matrix is ordered from $-\frac{M}{2}$ to $\frac{M}{2}$ for symmetry. Of course, adding a small translational shift, $\zeta$, to the magnetic vector potential $A=\langle-B(y+\zeta), 0,0\rangle$, would not change the magnetic field: $B=B\hat{z}$.

A final critical feature of these matrices is their translational invariance. That is, except for the varying effective potential, every one of the three matrices $\left[\mathbf{Z}-\mathbf{H}_{i,i}\right]$, $\mathbf{V}$, and $\mathbf{V^\dag}$, is identical no matter what slice $i$ is being calculated. This, of course, is logical given that (\ref{eq:discretized}) has an identical form for every lattice point and (\ref{NN}) has an identical form for every slice. In the presence of a non-translationally invariant potential, however, neither $\mathbf{H}_{i,i}$ nor the Green's function matrices themselves will be the same for every slice. This is an important result, because otherwise the Green's functions would provide no information about the energetic changes across the system.

\subsection{The Iterative Process}\label{sec:iterative}
To calculate the Green's functions across a sample an iterative process is used. First, an initial matrix, $\mathbf{G}_{0,0}$, is determined using boundary conditions (see section \ref{Bounds}), and then each ensuing Green's function, $\mathbf{G}_{1,1}, \mathbf{G}_{2,2}$ etc., is calculated from the previous Green's function. As explained earlier, this iterative process derives from the Dyson equation. For the purposes of the MacKinnon method, this equation should be recast:

\begin{equation}\label{DysonM}
\mathbf{G}^{(n+1)}_{i,j}=\mathbf{G}^{(n)}_{i,j}+ \mathbf{G}^{(n)}_{i,n}\mathbf{V}\mathbf{G}^{(n+1)}_{n+1,j} \textrm{  for  } (i,j\leq N)
\end{equation}
Where $i,j$, as before, represents the interaction between slices $i$ and $j$, and superscripts $n$ and $n+1$ represent the number of slices incorporated into the calculation thus far. Thus, for example, $\mathbf{G}_{2,3}^{(3)}$ is the Green's function representing the interaction between slices 2 and 3, calculated after iterating to slice 3. It is worth noting that this equation is very well-behaved upon repeated application. Numerous tests carried out during the writing of this thesis's local DOS program consistently showed that the Green's functions converged as iterations were carried out for systems of various lengths.

Generally, each of the applications of Green's functions require iterating from slice 0 to slice $N$. The goal is therefore to be able to obtain any Green's function in the system having iterated to slice $N$, since this takes into account all the energetics of the system. For example, $\mathbf{G}^{(N)}_{2, 3}$, is the Green's function linking slice 2 to slice 3 after the iteration has carried all the way through to the edge of the system at slice $N$. It is the definitive value for that interaction (leaving aside boundary conditions), as opposed to $\mathbf{G}^{(3)}_{2, 3}$, which is a matrix that does not take into account the impact on the slice 2-3 interaction resulting from the Hamiltonians of slices 4 all the way through $N$. $\mathbf{G}^{(N)}_{2, 3}$ differs from $\mathbf{G}^{(3)}_{2, 3}$ for non-zero field, because on each iteration, every $M \times M$ Green's function matrix to be calculated is multiplied by either $\mathbf{V}$ or $\mathbf{V^\dag}$, and these perturbation matrices equal the identity matrix only at $B$=0.

The details of carrying out this iterative process to slice $N$ for the density of states and transmission coefficients calculations require substantial specific explanations. Consequently, they are put off for Chapter Four. Here the general iterative method, moving on from the Dyson equation, is explicated.

The Dyson equation, though very useful, is recast into four equations derived from it \cite{MacDaddy}:

\begin{eqnarray}
\mathbf{G}^{(n+1)}_{n+1,n+1}&=&[\mathbf{Z}-\mathbf{H}_{n+1}-\mathbf{V^\dag} \mathbf{G}^{(n)}_{n,n}\mathbf{V}]^{-1} \label{FiveA}\\
\mathbf{G}^{(n+1)}_{i,j}&=&\mathbf{G}^{(n)}_{i,j}+\mathbf{G}^{(n)}_{i,n} 
\mathbf{V}\mathbf{G}^{(n+1)}_{n+1,n+1}\mathbf{V^\dag}\mathbf{G}^{(n)}_{n,j} \hspace{.8 cm} (i, j\leq N) \label{FiveB}\\
\mathbf{G}^{(n+1)}_{i,n+1}&=&\mathbf{G}^{(n)}_{i,n}\mathbf{V} \mathbf{G}^{(n+1)}_{n+1,n+1}\hspace{.8 cm}(i\leq N) \label{FiveC}\\
\mathbf{G}^{(n+1)}_{n+1,j}&=&\mathbf{G}^{(n+1)}_{n+1,n+1}\mathbf{V^\dag} \mathbf{G}^{(n)}_{n,j} \hspace{.8 cm} (j\leq N)\label{FiveD}
\end{eqnarray}

\begin{figure}
\centering
\includegraphics[width=0.9\textwidth, viewport= 70 60 640 350]{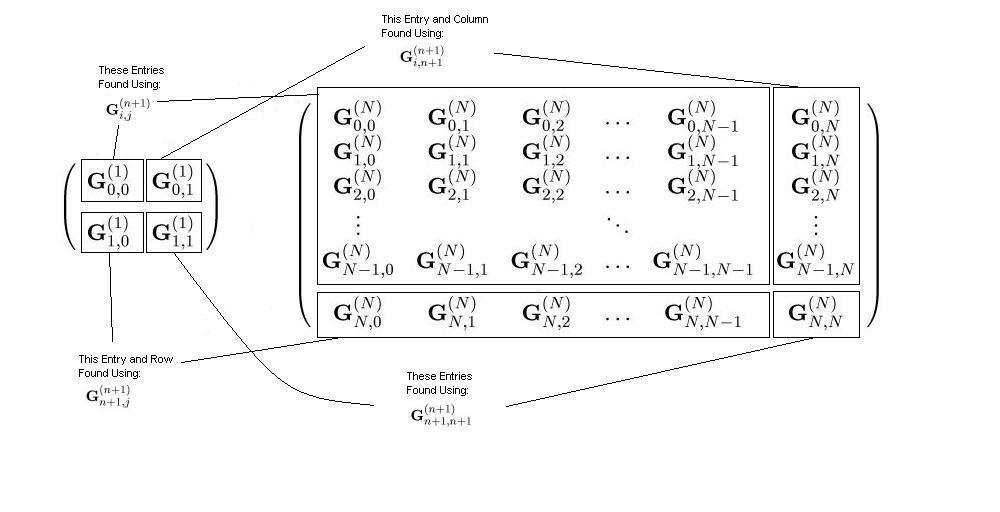}
\caption[Iterative Green's Functions]{\textit{The Range of the Four Iterative Green's Functions. The figure on the left describes which equations (of (\ref{FiveA}-\ref{FiveD})) are used to calculate the entries of $\mathbf{G}^{(1)}$ from $\mathbf{G}^{(0)}$. The right figure describes which equations are used to calculate the entries of $\mathbf{G}^{(N)}$ from $\mathbf{G}^{(N-1)}$. Since each Green's function matrix, $\mathbf{G}_{i,j}^{(N)}$, is an $M \times M$ matrix, this right figure represents the total Green's function matrix, $\mathbf{G}^{(N)}$, an $NM \times NM$ matrix. Only certain of the Green's function matrices within this total matrix are needed for most numerical applications; thus, the iterative method ends up saving a great deal of calculational time.}}\label{fig:GreensMap}
\end{figure}

These four equations divide up the total Green's matrix that takes into account $n+1$ slices, $\textbf{G}^{(n+1)}$, into four regions. The total Green's functions matrix at this iteration is an $n+1 \times n+1$ matrix with each entry itself being an $M \times M$ matrix. Each equation (\ref{FiveA})-(\ref{FiveD}) is capable of determining only certain $M \times M$ matrices in the total matrix, though together they can find them all (see Figure \ref{fig:GreensMap}).

Equation (\ref{FiveA}), finds the self-interaction energy of the $n+1$ slice, the last row, last column entry of total matrix after $n+1$ iterations, $\textbf{G}^{(n+1)}$. Equation (\ref{FiveB}) is capable of finding any entry in the total matrix except for the last row and the last column. It cannot find them all at once, however, and specific values for both $i$ and $j$ must be implemented in order for the iterative process to work. As will be shown in Chapter Four, this is not problematic, as the Green's functions for the density of states are almost always sought for the case of slice $i=j$ only. Equation (\ref{FiveC}) calculates the final $n+1$ column of the total matrix, while (\ref{FiveD}) finds the $n+1$ row of the total matrix. The derivation of each of these equations from the Dyson equation is offered in Appendix A. From these four matrix relations (\ref{FiveA})-(\ref{FiveD}) every Green's function relating any two slices of the quantum system can be found by iteration.

\subsection{Adding Leads: Green's Function Boundary Conditions} \label{Bounds}

The issue of the boundary conditions remains. Though the iterative process allows one to calculate every value of the Green's function across a system, it does not take into account the interactions of the system's edges with the leads. The most relevant boundary conditions to consider are where two semi-infinite metal leads are attached to each end of the quantum system (see Figure \ref{fig:twodchannel}). These conditions are imposed below and were those used in the numerical models of Chapter Four.

\subsubsection{Determining and Sorting Eigenvalues and Eigenvectors}
To begin applying boundary conditions, the matrix form of the Green's function equation (\ref{NN}) is recast as an eigenvalue problem:

\begin{equation}\label{InhomoM}
\left(\begin{array}{cc}
\mathbf{V} &0 \\
0& \mathbf{V} \\
\end{array}\right)
\left(\begin{array}{cc}
\mathbf{G}_{i+1,j} \\
\mathbf{V^\dag G}_{i,j} \end{array}\right) = 
\left(\begin{array}{cc}
\mathbf{Z-H} & \mathbf{-I}\\
\mathbf{I} & 0 \end{array}\right)
\left(\begin{array}{cc}
\mathbf{G}_{i,j} \\
\mathbf{V^\dag G}_{i-1,j} \end{array}\right)-
\left(\begin{array}{cc} 
\delta_{ij} \\
0 \end{array}\right)
\end{equation}
To solve for $\mathbf{G}$, the homogenous case of the same eigenvalue problem is considered. For eigenvalues $\alpha$ and eigenvector matrices $\mathbf{U^a}$ and $\mathbf{U^b}$ one obtains: 

\begin{equation}\label{HomoM}
\alpha
\left(\begin{array}{cc}
\mathbf{V} &0 \\
0& \mathbf{V} \\
\end{array}\right)
\left(\begin{array}{cc}
\mathbf{U}^a \\
\mathbf{U}^b \end{array}\right) =
\left(\begin{array}{cc} 
\mathbf{Z-H} & \mathbf{-I}\\
\mathbf{I} & 0 \end{array}\right)
\left(\begin{array}{cc}
\mathbf{U}^a \\
\mathbf{U}^b \end{array}\right)
\end{equation}

A single matrix whose eigensolutions are sought must be formed. To accomplish this, both sides are multiplied by the inverse of $\mathbf{V}$. Using the fact that $\mathbf{V}^{-1}=\mathbf{V^\dag}$, the result, known as the transfer matrix, is obtained:

\begin{equation}
\left(\begin{array}{cc}
\mathbf{V^\dag (Z-H)} & \mathbf{-V^\dag} \\
\mathbf{V^\dag}& 0 \\
\end{array}\right)
\end{equation}

The eigenvectors contained within  $\mathbf{U^a}$ represent the wavefunctions of the quantum system. They can be sorted according to the magnitude of their corresponding eigenvalues. If $\alpha > 1$, then the wavefunction is an evanescent mode traveling with positive momentum, while those vectors for which $\alpha < 1$ are evanescent modes traveling with negative momentum. In the case of $\alpha = 1$, the corresponding eigenvector is a conducting mode. These current carrying modes must be normalized.

Eigenvector $\mathbf{U^a}$ is separated into two $M\times M$ matrices, $\mathbf{U_+}$ and $\mathbf{U_-}$, corresponding to the direction of the wavefunctions' momenta.  Eigenvalues $\alpha$ are divided up appropriately into $\alpha_+$ (for $\alpha>1$ and positive momentum values of $\alpha=1$)  and $\alpha_-$ ($\alpha<1$ and negative momentum values of $\alpha=1$). The current-carrying modes, i.e. those wavefunctions with $\alpha=1$, are determined to have positive or negative momentum using the definition of current in a quantum system. If $2\psi^2_{n,m}  \textrm{Im}(\alpha \mathbf{V}_{m}) > 0$, then the current is positive (moves right). If this expression is $<0$, the current is negative (moves left).

\subsubsection{Applying Boundaries}
With  $\alpha_+$,  $\mathbf{U_+}$,  $\alpha_-$, and $\mathbf{U_-}$ in hand, an expression for the Green's functions corresponding to each boundary can be derived. $\mathbf{G}^{-\infty}_{0,0}$ is defined as the Green's function from the end of the semi-infinite lead on the left side of the sample to the zeroth slice of the sample. $\mathbf{G}^{+\infty}_{N,N}$ represents the Green's function from the final slice $N$ of the sample to the end of the semi-infinite lead on the right side of the sample.

The left lead, $\mathbf{G}^{-\infty}_{0,0}$, is considered first. The boundary condition here is that the Green's function relating slice $-\infty$ to slice $0$ must go to zero. The reason for this is that the Green's functions must decay into the lead:  the system's energy ought to go to zero as one moves infinitely far away from the quantum sample and into the current injector. From the comparison of the inhomogeneous and homogeneous eigenvalue formulations, (\ref{InhomoM}) and (\ref{HomoM}), and the fact that according to the Bloch theorem for a regular lattice, $\mathbf{G}_{i+1,j}=\alpha\mathbf{G}_{i,j}$, it is identified that: 

\begin{equation} \label{ZeroGreenBound}
\mathbf{G}^{-\infty}_{i,j}=\mathbf{U_+}\alpha^{i-j}_+\mathbf{A}
\end{equation}
for two arbitrary horizontal positions in the lead, $i$ and $j$, where $i \leq j$, and where $\mathbf{A}$ is a matrix of coefficients. $\vert \alpha \vert >1$ in order to satisfy the boundary condition (given that $i \leq j$). To derive an eigenvalue expression for $\mathbf{G}^{-\infty}_{i,j}$ independent of $\mathbf{A}$, two cases of different initial values $i=-1, j=0$ and  $i=0, j=0$ are considered and plugged into (\ref{InhomoM}).

\begin{eqnarray}
\mathbf{VG}_{0,0} & =& (\mathbf{Z} -\mathbf{H})\mathbf{G}_{-1,0}-\mathbf{VG}_{-2,0} -0 \\
\mathbf{VG}_{1,0} & =& (\mathbf{Z} -\mathbf{H})\mathbf{G}_{0,0}-\mathbf{VG}_{-1,0}-1
\end{eqnarray}

Immediately,  $\mathbf{VG}_{1,0}$ goes to zero because $i>j$. The values for $\mathbf{G}_{0,0}$, $\mathbf{G}_{-1,0}$, and $\mathbf{G}_{-2,0}$ are determined by use of (\ref{ZeroGreenBound}):

\begin{eqnarray}
\mathbf{G}_{0,0} & = & \mathbf{U_+}\mathbf{A} \\
\mathbf{G}_{-1,0}& = & \mathbf{U_+}\alpha^{-1}_+\mathbf{A} \\
\mathbf{G}_{-2,0}& = & \mathbf{U_+}\alpha^{-2}_+\mathbf{A}
\end{eqnarray}
Substituting these terms in and $(\mathbf{Z} -\mathbf{H})$ out, simplification leads to:

\begin{equation}
\mathbf{A}=\alpha^{-1}_+\mathbf{U_+}^{-1}\mathbf{V}^\dag
\end{equation}
And thus:

\begin{equation}
\mathbf{G}^{-\infty}_{0,0}=\mathbf{U_+}\alpha^{-1}_+\mathbf{U}_+^{-1}
\mathbf{V^\dag}
\end{equation}

A nearly identical process is applied to calculate the Green's function in the right lead. The Green's function again must decay into the lead, but for this to hold true here $i \geq j$ and $\vert \alpha \vert<1$. Comparing the inhomogeneous and homogeneous equations again one obtains: $\mathbf{G}^{+\infty}_{i,j}=\mathbf{U_-}\alpha^{j-i}_-\mathbf{A}$. Substituting in two sets of $i$ and $j$, the result is: 

\begin{equation}
\mathbf{G}^{+\infty}_{N,N}=\mathbf{U_-}\alpha_-\mathbf{U}_-^{-1}
\mathbf{V}
\end{equation}

The expressions $\mathbf{G}^{-\infty}_{0,0}$ and $\mathbf{G}^{+\infty}_{N,N}$ are inserted into the appropriate places in the numerical process. Carrying out the iterative process, one first inserts $\mathbf{G}^{-\infty}_{0,0}$ into the right-hand side of (\ref{FiveA}) in the place where $\mathbf{G}^{(0)}_{0,0}$ sits in (\ref{FiveA}). This will yield the result $\mathbf{G}^{(1)}_{1,1}$. The  $\mathbf{G}^{-\infty}_{0,0}$ expression is also inserted into (\ref{FiveC}) in the places where $\mathbf{G}^{(0)}_{i,0}$ sits, and is inserted into (\ref{FiveD}) in the place where $\mathbf{G}^{(0)}_{0,j}$ sits. Applying this expression in those cases gives the Green's functions, $\mathbf{G}^{(1)}_{0,1}$ and $\mathbf{G}^{(1)}_{1,0}$, respectively. Furthermore,  $\mathbf{G}^{-\infty}_{0,0}$ must be inserted into (\ref{FiveB}) on the first iteration in place of $\mathbf{G}^{(0)}_{i,j}$, $\mathbf{G}^{(0)}_{i,0}$ , and $\mathbf{G}^{(0)}_{0,j}$, returning, $\mathbf{G}^{(1)}_{0,0}$. Applying  $\mathbf{G}^{-\infty}_{0,0}$ into these four equations on the first iteration, allows one to take into account the Green's functions running all the way into the left lead.

The implementation of the right lead has one slight nuance. It is inserted only into (\ref{FiveA}). Since (\ref{FiveB})-(\ref{FiveD}) each depend upon the $\mathbf{G}^{(n+1)}_{n+1,n+1}$ matrix emerging from (\ref{FiveA}), implementing the right lead Green's function once in (\ref{FiveA}) is sufficient. The definition of the Green's function as $(\mathbf{Z-H}_{n+1})^{-1}$ is used and $(\mathbf{G}^{+\infty}_{N,N})^{-1}$ is inserted into its place:

\begin{equation}
\mathbf{G}^{+\infty}_{N+1,N+1}=[(\mathbf{G}^{+\infty}_{N,N})^{-1}
-\mathbf{V^\dag} \mathbf{G}^{N}_{N,N}\mathbf{V}]^{-1}
\end{equation}

This completes the formalism of the Green's function numerical model. Iterating Green's functions provides a powerful and efficient tool for calculating fundamental quantum mechanical properties of electronic systems. This is evinced in Chapter Four, where numerical Green's functions are applied to calculating the local density of states and transmission properties of one-dimensional quantum samples.

\chapter{The Density of States Model}

\section{Introduction}
In this chapter, two programs for calculating properties of quantum systems are presented and their results are analyzed. The first program calculates the local Density of States (DOS) of a 1D quantum sample, the second the transmission coefficients of a 1D quantum sample. The local DOS is a fundamental property of a quantum system and the use of numerical Green's functions in a discretized lattice provide an effective probe of its behavior. The two programs were written in Visual C\# (C-Sharp), using CenterSpace.Matrix auxiliary code and NPlot Graphics code. All code developed is the work of the author with two exceptions: a piece of code that made assigning eigenvectors to positive and negative momenta matrices more efficient, and a piece of code in the transmission program that rapidly carried out a series of multiplications for the transmission and reflection coefficients. These two pieces of code were based on the work of C. H. W. Barnes \cite{BarnesCode}.

This chapter will begin by discussing the local DOS and the method by which the program was compiled. It will then compare results generated from the numerical model with expected theoretical results and will confirm the model's very high degree of accuracy. Finally, the method of the transmission program will be presented and its results analyzed.

\section{The Density of States Program}
\subsection{The Density of States Function}

The density of states is a function of energy that measures the number of states available in a given energy range per unit length. The number of available states depends upon the number of occupiable states in k-space, and the DOS is in general a measure of how closely packed energy levels (and their corresponding wavefunctions) are in a quantum system. The one-dimensional total density of states is given by:

\begin{equation}\label{Blahblah}
\rho(E)=\frac{-1}{\pi NM}  \textrm{Im} \left(\sum_{i=1}^{N}Tr(\mathbf{G}^{(N)}_{i,i}) 
\right)
\end{equation}
where, as before, $N$ is the length of the 1D system in lattice units, $M$ is its width in lattice units, and $i$ is a slice of the system (see Fig. \ref{fig:LatticeMap}). This equation (\ref{Blahblah}) is the density of states averaged out across the sample. To determine the local density of states, the density of states is individually measured at every lattice point $m$ on slice $i$, and the relation to the Green's functions becomes: 

\begin{equation} \label{GreenDOS}
\rho(m)=\frac{-1}{\pi}  \textrm{Im} \left(\mathbf{G}^{(N)}_{i,i}\right)
\end{equation}
evaluated at entry $(m,m)$ in the $\mathbf{G}^{(N)}_{i,i}$ matrix.

This result emerges from an elegant physical and mathematical argument highlighting the fundamental nature of the DOS. The density per energy value $\epsilon$ is $\sum_n\delta(\epsilon-\epsilon_n)$, where $\epsilon_n$ is an eigenvalue of the Hamiltonian. The DOS then is the product of this density per energy level and the corresponding wavefunction probability map, $\vert\psi(r)^2 \vert$:

\begin{equation}\label{DOSDelta}
\rho(r)= \sum_n \delta(\epsilon -\epsilon_n)\vert\psi(r)^2 \vert 
\end{equation}

To link this to Green's functions, one must return to the original Green's function formalism. Due to the poles in the Green's function in (\ref{En}), a branch cut of $G(r, r', z)$ is taken along the real-axis, creating two new Green's functions, the retarded and advanced Green's functions: $G^+= \lim_{s \rightarrow 0^+}G(\epsilon + is)$ and $G^-= \lim_{s \rightarrow 0^+}G(\epsilon - is)$. The retarded Green's function, $z=\epsilon +is$, was selected for use in the calculation. Using the identity:

\begin {equation}
\lim_{y \rightarrow 0^+} \frac{1}{x \pm iy}= P(\frac{1}{x}) \mp i\pi\delta(x)
\end{equation}
where $P$ is the Principle Value Term, one can recast the retarded Green's function. Defining $x=\epsilon-\epsilon_n$ and $y=s$ (since $\frac {1}{z-\epsilon_n}=\frac{1}{\epsilon-\epsilon_n+is}$) one arrives at:

\begin{equation}\label{GreenP}
G^+(r,r', \epsilon) = P \sum_n \frac{\psi_n(r) \psi_n^*(r')}{\epsilon- \epsilon_n} - i\pi \sum_n \delta(\epsilon -\epsilon_n)\psi_n(r) \psi_n^*(r')
\end{equation}

And therefore:

\begin{equation}\label{DaDOS}
\rho(r)= - \frac{1}{\pi}  \textrm{Im} \left(G^+ (r, r', \epsilon)\right)
\end{equation}

This demonstrates the very close link between Green's functions and the DOS. By calculating the Green's function at every $\psi_{n,m}$ in the lattice, the density of states is found. No simplifications or approximations whatsoever (except to create the discrete lattice) need be made.

\subsection{The Method of the Program}

In creating the program, most of the input constants, variables, and matrices are simply defined and plugged in to meet the specifications of the system. The one area that is tricky, and which this section will spend substantial time dealing with is the iteration to the proper slice in order to calculate the local DOS. 

To begin, the lattice width $M$, lattice length $N$, lattice spacing constant $a$, and applied perpendicular magnetic field $B$ are defined. Each of these variables can be readily varied. The effective mass used was that of GaAs, $0.067e$. The constants $\nu=-\frac{2ma^2}{\hbar^2}V$ and $\epsilon= 4-\frac{2ma^2}{\hbar^2}E$ as described before are used, where $E$ is the input voltage in meV. Infinitely small complex value $is$ is added to $\epsilon$. Matrices $\mathbf{V}$, $\mathbf{V}^\dag$, and $\mathbf{Z}$ are initialized with appropriate values, as explained in Chapter Three.

The Hamiltonian matrix is an effective potential matrix with -1 on the off diagonals. The effective potential values $\nu$ run down the main diagonal. Recall that this matrix represents the Hamiltonian for a slice of the system only. For the case a translationally-invariant potential (e.g. infinite square well or harmonic oscillator), every slice will have the same Hamiltonian, each main diagonal entry corresponding to the potential at a lattice point as a function of $m$. For example, the first matrix entry corresponds to lattice point $m=-M/2$ at the bottom edge of the system, and therefore must reflect the potential---including the effect of the magnetic field---a distance $M/2$ from the system's center.

Specifically, for the case of an infinite square well, $\nu=0$ all the way down the diagonal, while for a harmonic oscillator potential, $V=\frac{\omega}{2}(m-\frac{M-1}{2})^2$, giving a minimum at matrix entry, $M/2$, which corresponds to the center of the system, row $m=0$ (see Fig. \ref{fig:LatticeMap}). For the more complicated translationally-varying potential, like that emerging from a surface gate, the matrix must also be made a function of $N$, and it must be recalculated at every iteration.

With the essential matrices in hand, the eigenvalues and eigenvectors of the transfer matrix are found and sorted by modulus. This is a numerically-direct but programatically-heavy technique. In short, code was written to effectively sort the $N$ eigenvectors and eigenvalues by their modulus and keep the eigenvectors and eigenvalues appropriately paired. A separate piece of code deals with those eigenvalues whose modulus is equal to one---the current-carrying modes---and sorts them into positive and negative current. Still another piece of code sorts the $\alpha>1$ and positive momentum values of $\alpha=1$ into $\alpha_+$ matrix and the $\alpha<1$ and negative momentum values of $\alpha=1$ into $\alpha_-$. After $\alpha_+$, $\alpha_-$, $\mathbf{U_+}$, and $\mathbf{U_-}$ have been found and sorted, and after they are used to determine $\mathbf{G}^{-\infty}_{0,0}$, the iterative process begins.

Looking back to (\ref{GreenDOS}), one sees that in order to calculate the density of states iteratively, one must find the $\mathbf{G}^{(N)}_{i,i}$ matrix. This is the Green's function self-interaction at slice $i$, analogous to $G^+(r,r', \epsilon)$ of (\ref{GreenP}) with $r=r'$. Recall that the superscript in $\mathbf{G}^{(N)}_{i,i}$ indicates this Green's function is not the result of iterating to slice $i$ (which would be $\mathbf{G}^{(i)}_{i,i}$), but is an iteration all the way from slice 0 to $N$. In fact, since boundary conditions are included, it is an iteration from $-\infty$ to $\infty$. Therefore, to calculate the local density of states using the iterative method, a process is needed to calculate the Green's functions at any slice inside the system while accounting for an iteration all the way to $\infty$.

This method will naturally rely upon (\ref{FiveB}), with $i=j$ and $n=N$; however, finding this value depends upon iterating the other equations in the proper order. The method developed is as follows. The program calculates (\ref{FiveA}) from slice 0 (using $\mathbf{G}^{-\infty}_{0,0}$ as a starting point as explained in \ref{Bounds}) until a particular slice $i=w$. It is at slice $w$ where the local density of states will be calculated. Note, again, that when calculating (\ref{FiveA}), if the potential varies with $N$, the Hamiltonian must be recalculated at every iteration, thereby slightly increasing calculation time.

Once the program has iterated to slice $w$, the value of $\mathbf{G}^{(w)}_{w,w}$ is stored and then (\ref{FiveA}) is applied again. Then (\ref{FiveB}) is applied, using the stored input of $\mathbf{G}^{(w)}_{w,w}$ three times and $\mathbf{G}^{(w+1)}_{w+1,w+1}$ to obtain $\mathbf{G}^{(w+1)}_{w,w}$. Subsequently,  (\ref{FiveC}) and  (\ref{FiveD}) are applied, employing $\mathbf{G}^w_{w,w}$ and $\mathbf{G}^{(w+1)}_{w+1,w+1}$ once each to calculate $\mathbf{G}^{(w+1)}_{w,w+1}$ and $\mathbf{G}^{(w+1)}_{w+1,w}$ respectively. In the next iteration, applying (\ref{FiveA}) will yield $\mathbf{G}^{(w+2)}_{w+2,w+2}$. When iterating (\ref{FiveB}) again, it will be given by:

\begin{equation}
\mathbf{G}^{(w+2)}_{w,w}= \mathbf{G}^{(w+1)}_{w,w}+\mathbf{G}^{(w+1)}_{w,w+1}\mathbf{V}\mathbf{G}^{(w+2)}_{w+2,w+2}\mathbf{V}^\dag\mathbf{G}^{(w+1)}_{w+1,w}
\end{equation}

This process explains why finding $\mathbf{G}^{(w+2)}_{w,w}$ depends upon the other iterative relations. It requires (\ref{FiveA}) and its $\mathbf{G}^{(w+2)}_{w+2,w+2}$ result, (\ref{FiveC}) and its $\mathbf{G}^{(w+1)}_{w,w+1}$ result, (\ref{FiveD}) and its $\mathbf{G}^{(w+1)}_{w+1,w}$ result, and (\ref{FiveB}) itself and its $\mathbf{G}^{(w+1)}_{w,w}$ result. Equation (\ref{FiveB}) must be iterated after (\ref{FiveA}), but before (\ref{FiveC}), and (\ref{FiveD}), since (\ref{FiveB}) relies upon their values from the previous iteration. These four equations are then applied over and over again until slice $N$, at which point the desired $\mathbf{G}^{(N)}_{w,w}$ matrix is obtained.

At this point the Green's function in the right lead, $\mathbf{G}^{+\infty}_{N,N}$, is inserted into (\ref{FiveA}), according to the procedure described previously; then this result is used to find (\ref{FiveB}). This yields the ultimate result: $\mathbf{G}^{+\infty}_{w,w}$, which can also be thought of as $\mathbf{G}^{(-\infty \rightarrow +\infty)}_{w,w}$, since it takes into account the Green's functions in the leads and in every slice of the sample in between, running from $-\infty$ to $+\infty$. Each diagonal entry in  $\mathbf{G}^{+\infty}_{w,w}$ is the value of the Green's function at lattice point $m$, beginning from the bottom of the lattice slice and moving to the top. Taking the imaginary portion of this Green's function according to (\ref{DaDOS}) yields the local DOS at that lattice point alone. In order to calculate the density of states in every slice of the sample, the code found the local DOS for slice $w=1$, then started over again and calculated it for slice $w=2$, repeating the process until slice $N$. Thus the local density of states across the entire 1D system, in both $x$ and $y$, was determined. The code stored every lattice point's local DOS in a two-dimensional array where each column corresponded to a slice; these values were then fed into a graphics compiler.

Given this method, the degree to which $N$ is increased or decreased will lengthen or shorten the calculation time by the fractional change in $N$ squared. For example, the system doubled in length from 200 to 400 lattice slices, each iteration of a slice would require iterating twice as far to get to $N$ and then the program would also require twice as many iterations in order to include all $N$ slices. There is, unfortunately, no way around this fact, as the density of states not only must be calculated at each slice but also must take into account---on every iteration---the Green's function linking every slice to its neighbors. A typical calculation, with $M=41$ and $N=200$, took five minutes on a standard desktop computer.

\section{Results from the Density of States Program}

\subsection{Testing the program's accuracy}
The Green's function numerical model developed for this thesis matched expected results for the local DOS with extreme precision. First, the shape and response of the wavefunctions to varying magnetic field were examined. Next, the properties and behavior of the real bands of the system at varying magnetic field were considered. Lastly, Green's functions plots for a slice of the local density of states were compared with results using the analytic formula for the DOS derived from the carrier density. In all three cases, the Green's function model demonstrated great precision and reliability.

\subsubsection{Wavefunctions}
The first check of the system's dependability was an analysis of the lowest order wavefunction of the system, the first subband. For an infinite square well effective potential, the analytic solutions go like $\sin(nx)$ where $n$ is an integer corresponding to the subband number. 

Solving for the wavefunction is not a trivial task, however, as one must pick it out from among the $2M$ randomly sorted, eigenvectors, $\mathbf{U_+}$ and $\mathbf{U_-}$ in the system. Fortunately, the code had already sorted the eigenvectors into conducting and non-conducting modes. Since only the conducting modes' eigenvectors represented the real solutions to the Schr\"{o}dinger equation, only the current-carrying modes needed to be sifted. 

This sorting of the current carrying modes was achieved using the relation $e^{ik}=\alpha$. This derives from the translational invariance of the Bloch Function:

\begin{equation}
e^{ik}
\left(\begin{array}{cc} 
\psi_n  \\
\psi_{n-1} \\
\end{array}\right)
=\left(\begin{array}{cc}
\psi_{n+1}  \\
\psi_{n} \\
\end{array}\right)
\end{equation}
where $\psi_n$ is wavefunction one lattice point to the left of $\psi_{n+1}$. Since an infinite square well potential has a translationally-invariant potential in the x-direction, the eigenvalues $\alpha$ that solve the homogeneous equation (\ref{HomoM}), are identified as $e^{ik}$. 

The largest value of $k$ corresponds to the lowest order wavefunction, because the dispersion relation for each subband is a parabola, and the subbands are placed one above the other. Thus, for a given intersection of $E$, the Fermi energy, with the dispersion curves, the highest value of $k$ will be on the lowest energy subband. Once the lowest order wavefunction for both positive and negative momentum is identified, they can be plotted across the width of the sample (i.e. in the $y$ direction).

\begin{figure}
\centering
\includegraphics[width=0.9\textwidth, viewport= 30 60 870 330]{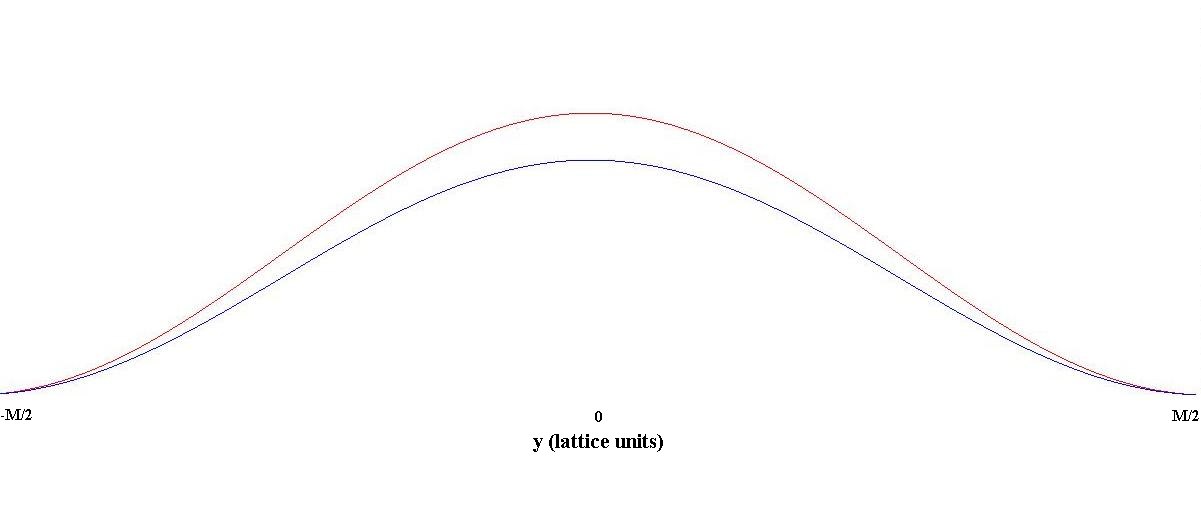}
\caption[Wavefunctions at $B$=0]{\textit{Wavefunctions generated with numerical Green's function code at zero magnetic field. Red (upper trace) represents the lowest-energy wavefunction moving to the right, blue (lower trace) the lowest-energy wavefunction moving left. The plots run across a slice, from $y$=-M/2 on the left to $y$=M/2 on the right. The positive wavefunction is scaled up so that the wavefunctions do not sit right atop one another.}}\label{fig:ZeroOrderPsis}
\end{figure}

\begin{figure}
\centering
\includegraphics[width=0.9\textwidth, viewport= 30 40 870 290]{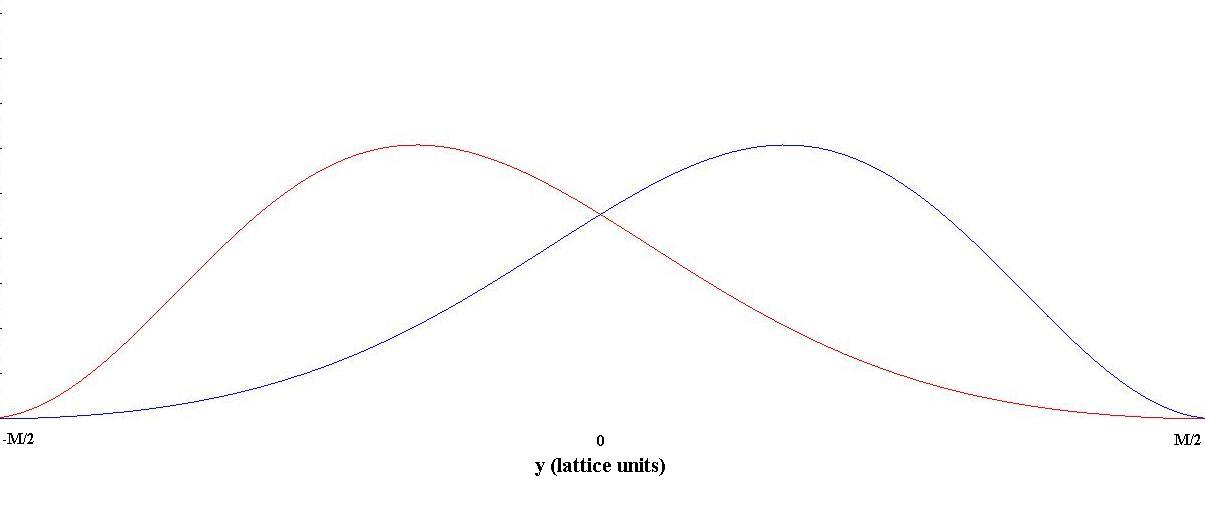}
\caption[Wavefunctions at $B$=0.1]{\textit{Wavefunctions at $B$=0.1 T. The Lorentz force pushes the wavefunctions moving in opposite directions against opposite walls.}}\label{fig:ZeroOrderPsisB01}
\end{figure}

\begin{figure}
\centering
\includegraphics[width=0.9\textwidth, viewport= 30 40 870 260]{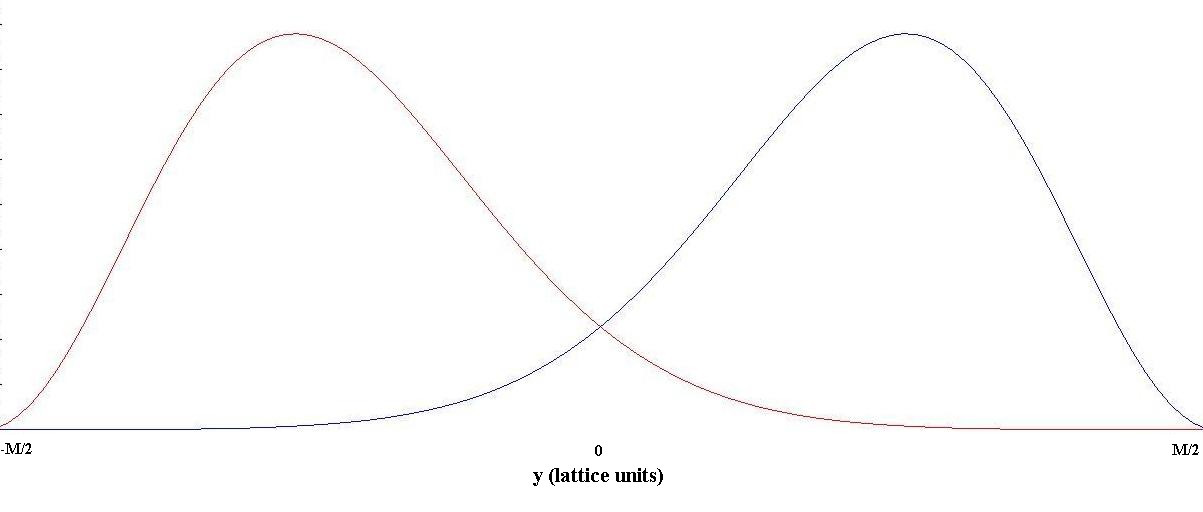}
\caption[Wavefunctions at $B$=0.3]{\textit{Wavefunctions at $B$=0.3 T.}}\label{fig:ZeroOrderPsisB03}
\end{figure}

\begin{figure}
\centering
\includegraphics[width=0.9\textwidth, viewport= 30 20 870 550]{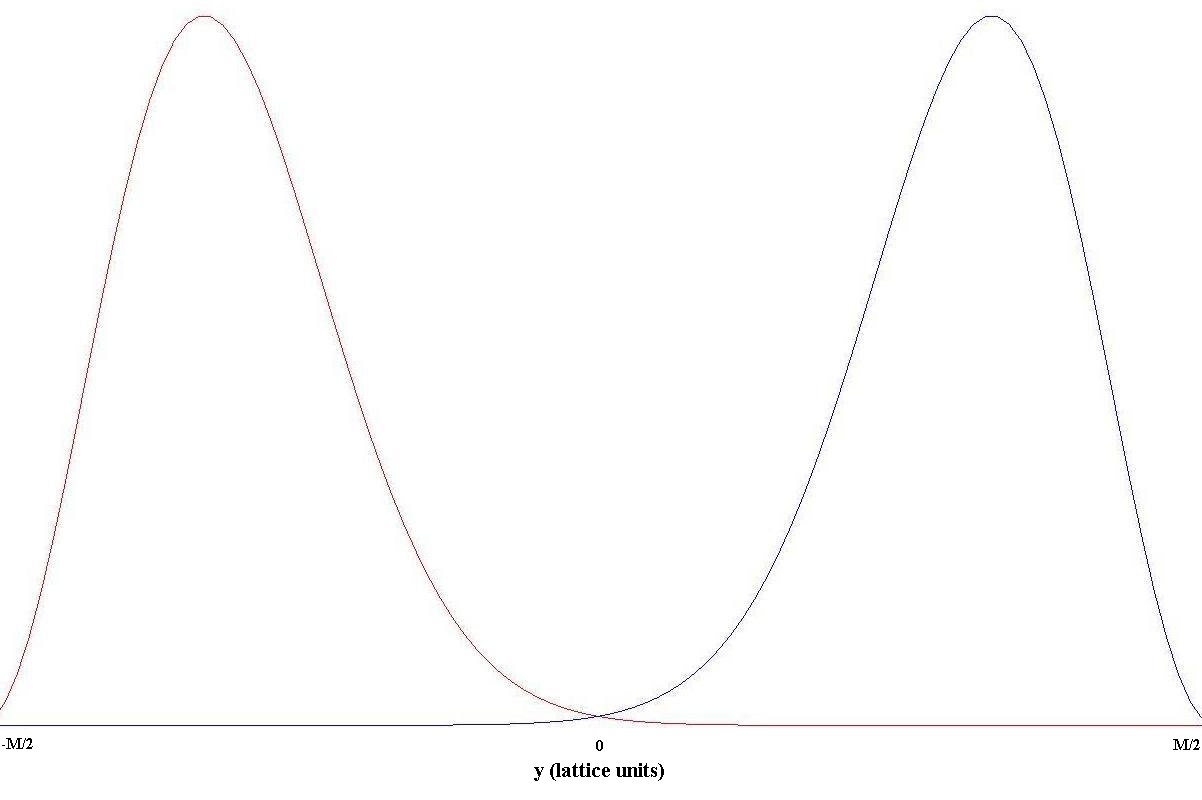}
\caption[Wavefunctions at $B$=1]{\textit{Wavefunctions at $B$=1 T.}}\label{fig:ZeroOrderPsisB1}
\end{figure}

At zero magnetic field, these two lowest order wavefunctions, simple $\sin^2(x)$ waves (since these are the lowest order eigenvector solutions under an infinite square well potential), should be indistinguishable, and indeed they are (see Figure \ref{fig:ZeroOrderPsis} where the positive momentum wavefunction is scaled up by a factor of 1.1 for clarity). As the magnetic field is increased, it is expected that the Lorentz force should begin to impact the wavefunctions, pushing them to opposite ends of the channel (since they move in opposite directions). This splitting is precisely what was found (see Figures \ref{fig:ZeroOrderPsisB01}-\ref{fig:ZeroOrderPsisB1}). Higher values of magnetic field meant the wavefunctions were pushed further towards the edges of the system (compare Fig. \ref{fig:ZeroOrderPsisB03} with Fig. \ref{fig:ZeroOrderPsisB1}). 

\subsubsection{Real Bands}
The next test conducted to make certain the system, particularly the eigenvalues and eigenvectors, behaved as expected was to look at the real band dispersion relations of the system. Since the value of $k$ (specifically $k_x$) is buried inside the wavefunction, one method to extract its value is to take advantage of the Bloch function's result, $e^{ik}=\alpha$. 

As mentioned above, the dispersion relation between $E$ and $k$ is parabolic for a 1D system. This is because solving the Schr\"{o}dinger Equation for this situation one obtains: 

\begin{equation}\label{SchroSubEqn}
E_{k_x,n}=E_{0,n}+\frac{\hbar^2k_x^2}{2m^*}
\end{equation}
where $E_{0,n}=(n+\frac{1}{2})\frac{\hbar eB}{m}$ is the subband energy.

At $B=0$, one therefore expects plots of each subband to look parabolic. Furthermore, for an infinite square well potential, it is well known that the solutions are spaced apart like $n^2$, and hence at $B=0$, one expects the parabolically shaped subbands to be spaced increasingly far apart. The numerical Green's Function model confirmed these expected results.

For a system with non-zero, perpendicular magnetic field given by the Landau gauge, however, the Lorentz force creates an effective confining potential given by: 

\begin{equation}\label{HarmOsc}
V_B(y)=\frac{1}{2}m^*\omega_c^2(y-y_0)^2 \hspace{.8 cm} \textrm{where }\hspace{.3 cm}  \omega_c=\frac{eB}{m^*}\hspace{.2 cm} \textrm{ \& } \hspace{.2 cm} y_0=\frac{\hbar k_x}{eB}
\end{equation} 
As the magnetic field increases, the confinement due to this potential increases. As $B$ rises, the magnetic confinement potential will have a greater impact on the subband energies than the infinite square well potential will. This is especially true for the lower bands, as the potential arising from the magnetic field is subband independent, while the higher the subband number the greater the subband energy due to the square well potential. 

Since the energy spacings resulting from a harmonic oscillator potential (\ref{HarmOsc}) are evenly separated, one would expect to see the spacing between subbands change from increasing spacing (like $n^2$) to even spacing as magnetic field rises. Moreover, the spacing between lower order subbands should become even first. This result was exactly what was found using the numerical Green's function model to plot real bands at various magnetic fields.

An additional feature worth confirming is the shape of these dispersion curves. At $B=0$, one expects to see strictly parabolic-shaped bands. That is, if $k$ is made to be non-zero then the energy must increase parabolically. This is not the case for $B\neq 0$. In this scenario, because the cyclotron orbits fit within the square well potential boundaries, even if $k$ is increased, the energy does not necessarily increase. That is, the energies become dispersionless once $B$ is sufficiently large. Therefore, the plots of the real bands become flat bottomed, not parabolic shaped, at sufficiently large $B$. The program again confirmed these expected results.

\subsubsection{Comparison with Analytic Evaluation}

As a third test of the model, local DOS plots using the Green's function code were compared to plots generated using the analytic solutions for the local DOS. For a one-dimensional quantum system, the system's carrier density is given by: 

\begin{equation}\label{CarrDens}
n_n=\frac{1}{\pi\hbar}\sqrt{2m^*(\epsilon-\epsilon_{0,n})} \hspace{.8 cm} \textrm{ for } \epsilon_{0,n}<\epsilon
\end{equation}
where $\epsilon_{0,n}$ is the subband energy. The density of states across a slice, $\rho(m)$, can be given not only by the method derived above, but also by taking the derivative of the carrier density of each subband multiplied by the probability map of each wavefunction: 

\begin{equation}\label{explicitDOS}
\rho(m)=|\psi_1^2|\frac{dn_1}{d\epsilon} + |\psi_2^2|\frac{dn_2}{d\epsilon}+
\ldots+ |\psi_p^2|\frac{dn_p}{d\epsilon}
\end{equation}
where $p$ is the number of conducting subbands. This equation can be seen as equivalent to the general definition of the density of states presented in (\ref{DOSDelta}) :

\begin{equation}
\rho(E)=\sum_n\delta(\epsilon-\epsilon_n)\vert\psi ^2 \vert
\end{equation} 

Given the parameters of the eigenvectors, the DOS, for positive momentum only, is equivalent to:

\begin{equation}\label{UPlusD}
\rho(m)=\frac{2m^*}{\pi\hbar}\sum_{l=0}^{p} \frac{\mathbf{U}_+^2[m,l]} 
{\sqrt{\epsilon-(l+\frac{1}{2})\frac{\hbar eB}{2m^*}}}
\end{equation}
where $m^*$ should not be confused with the lattice position, $m$, and $\mathbf{U}_+^2[m,l]$ is the probability of finding the $l$th conducting positive eigenvector at lattice point $m$ on any slice (assuming a translationally-invariant potential). Equation (\ref{UPlusD}) is only half the summation for local density of states at point $m$, as $\mathbf{U}_-^2[l,m]$ must also be taken into account. By calculating $\rho(m)$ at every lattice point on a slice, it can be plotted as a function of $m$ across the slice.

As both $\epsilon$ and $B$ are varied, it is found that the analytical expression and the Green's functions provide identical plots. This correspondence between values for the local density of states found using Green's functions and the local DOS calculated using an exact solution, evince yet again that numerical Green's functions derived from a tight-binding Hamiltonian provide an extremely accurate model. The specific behavior of the local density of states along a slice is analyzed in the following two sections. 

\subsection{Behavior of the Density of States at Zero Magnetic Field}

\begin{figure}
\centering
\includegraphics[width=0.9\textwidth, viewport= 60 280 790 730]{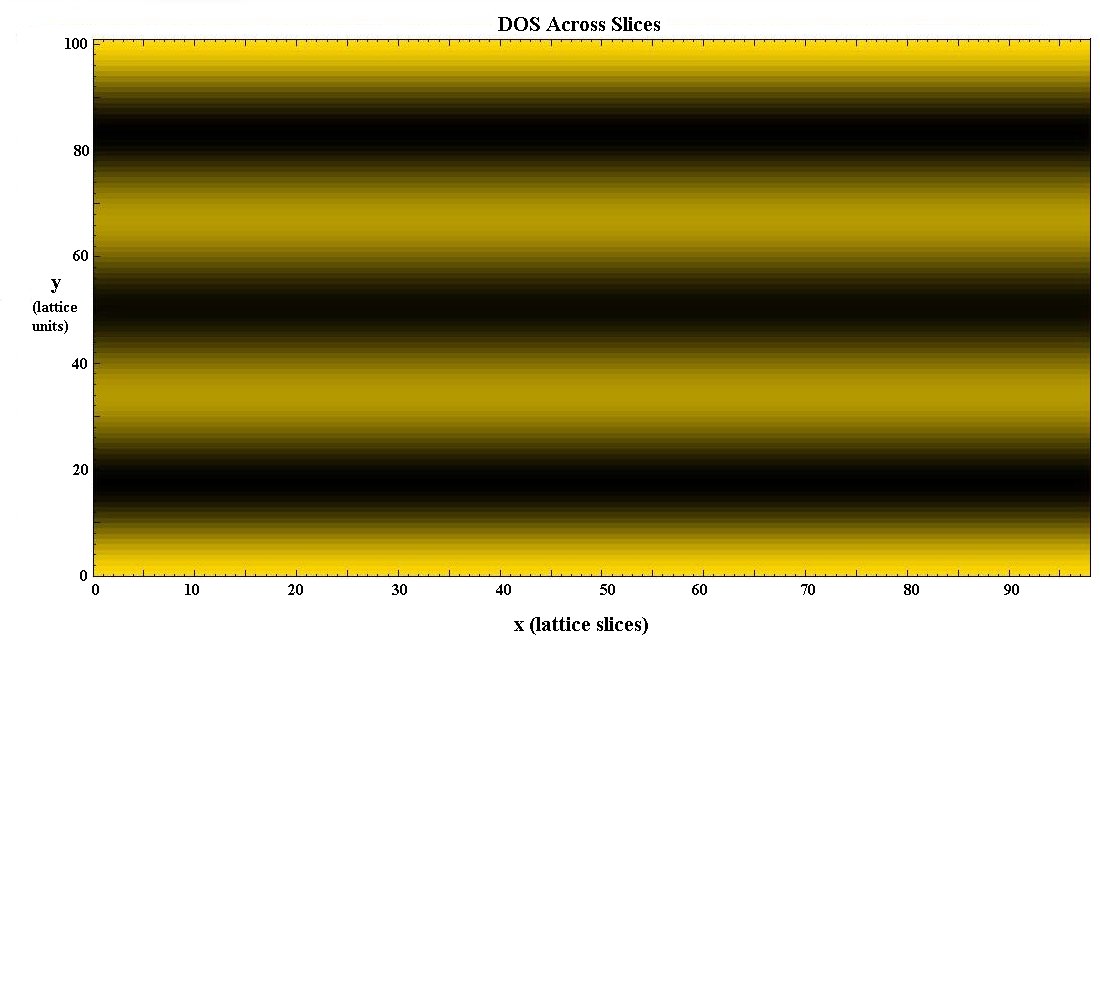}
\caption[2-D DOS in Zero Field]{\textit{Local Density of States in two-dimensions for an energy of three subbands at zero field. Black regions represent location of high DOS, yellow low DOS. Calculated using numerical Green's functions model. Note the perfect translational symmetry because of the translationally-invariant potential used.}}\label{fig:TwoDStraight}
\end{figure}

So long as the potential is translationally invariant in the x-direction, then the density of states plots across a slice should be identical for every slice. When surface gates are attached to the sample, however, this translational invariance is broken and the
density of states will vary for different slices. Therefore, without gates, it does
not matter which slice is considered. This is confirmed both by 2-D plots 
(see Figure \ref{fig:TwoDStraight}) and by mapping the DOS across a slice for many
slices and subtracting them to show they are in fact identical.

\begin{figure}
\centering
\includegraphics[width=0.9\textwidth, viewport= 30 170 820 760]{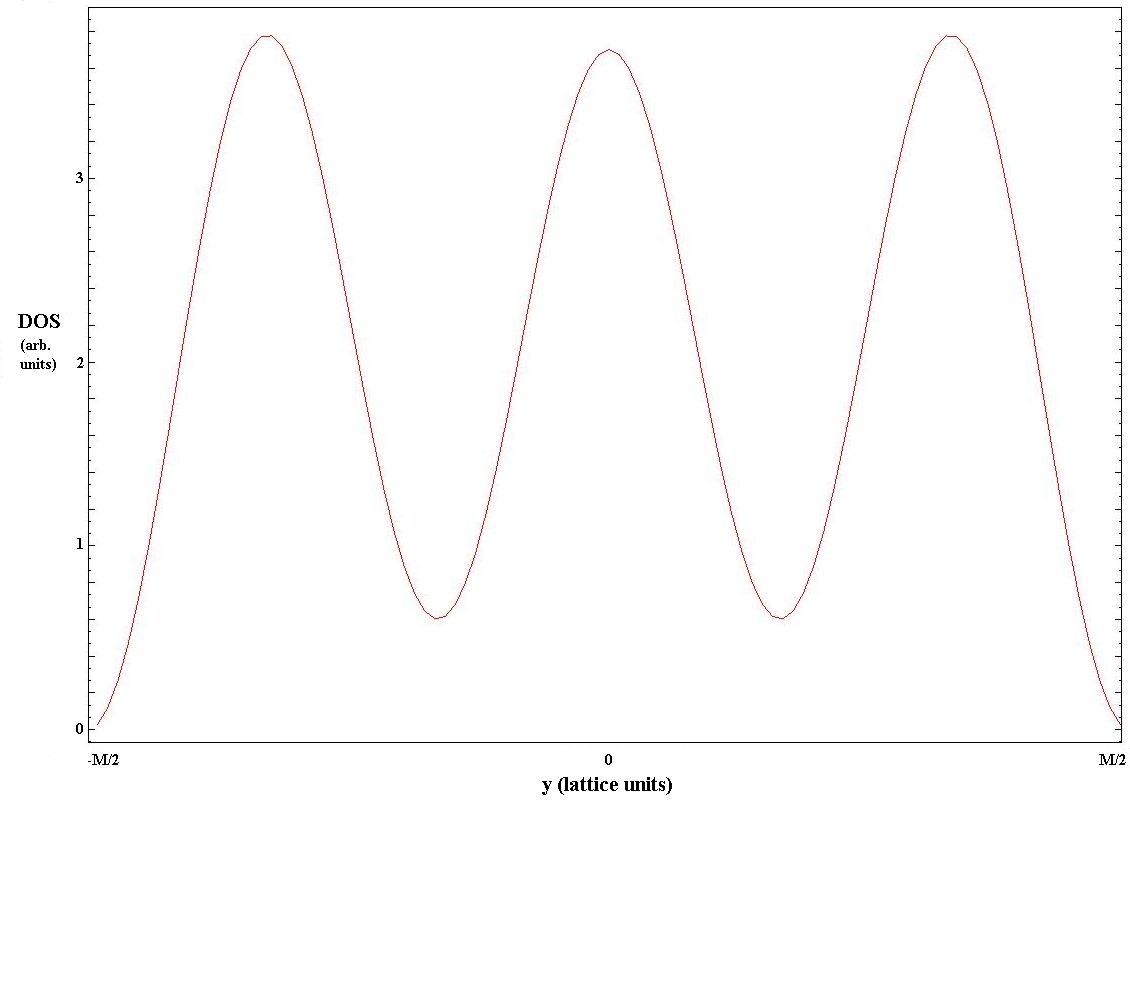}
\caption[DOS just above a subband]{\textit{Local DOS at Fermi energy just greater than that of the third subband. Generated using numerical Green's Functions.}}\label{fig:ThreePeaksNS}
\end{figure}

Considering first the case of no applied magnetic field, the system is examined as the Fermi energy increases. One observes that as $E$ grows, the number of bands grows as well. This is to be expected: at a higher Fermi energy there ought to be more available eigenenergies. One further sees that at the point where $E$ attains a value just greater than an eigenenergy---and hence a band appears---the density of 
states across the sample becomes very large. This
result is exactly as is to be expected for a 1DEG given that the DOS is the derivative of the
carrier density (see (\ref{CarrDens})): 

\begin{equation}
\rho(E)=2\sum_{n, E_{0,n}\leq E}\frac{\sqrt{2m^*}}{2\pi\hbar}(E-E_{0,n})^{-\frac{1}{2}}
\end{equation}
Given the form of this equation, there should be sharp peaks in $\rho(E)$ where $E-E_{0,n}$ is a minimum, that is, where the subband energy is just less than the Fermi energy $E$. This is precisely what was found (see Figure \ref{fig:ThreePeaksNS})

The physical reason underlying this peak behavior results from the density of states being directly related to the sum of the wavefunctions squared (\ref{explicitDOS}). These DOS plots should exhibit the shape of the highest subband's wavefunction squared, since the highest subband will have the highest value of $E_{0,n}$ and thus the highest value of $(E-E_{0,n})^{-\frac{1}{2}}$ and therefore dominates. Thus if, for example, the third subband's energy, $E_{0,3}$ is just less than $E$, there should be three sharp peaks with the troughs very near zero. The plots using Green's functions confirm this expectation (Fig. \ref{fig:ThreePeaksNS}). Since no magnetic field is present in this case, the positive and negative momenta wavefunctions overlap. This is not true under the presence of a  finite magnetic field. 

\begin{figure}
\centering
\includegraphics[width=0.9\textwidth, viewport= 40 260 830 530]{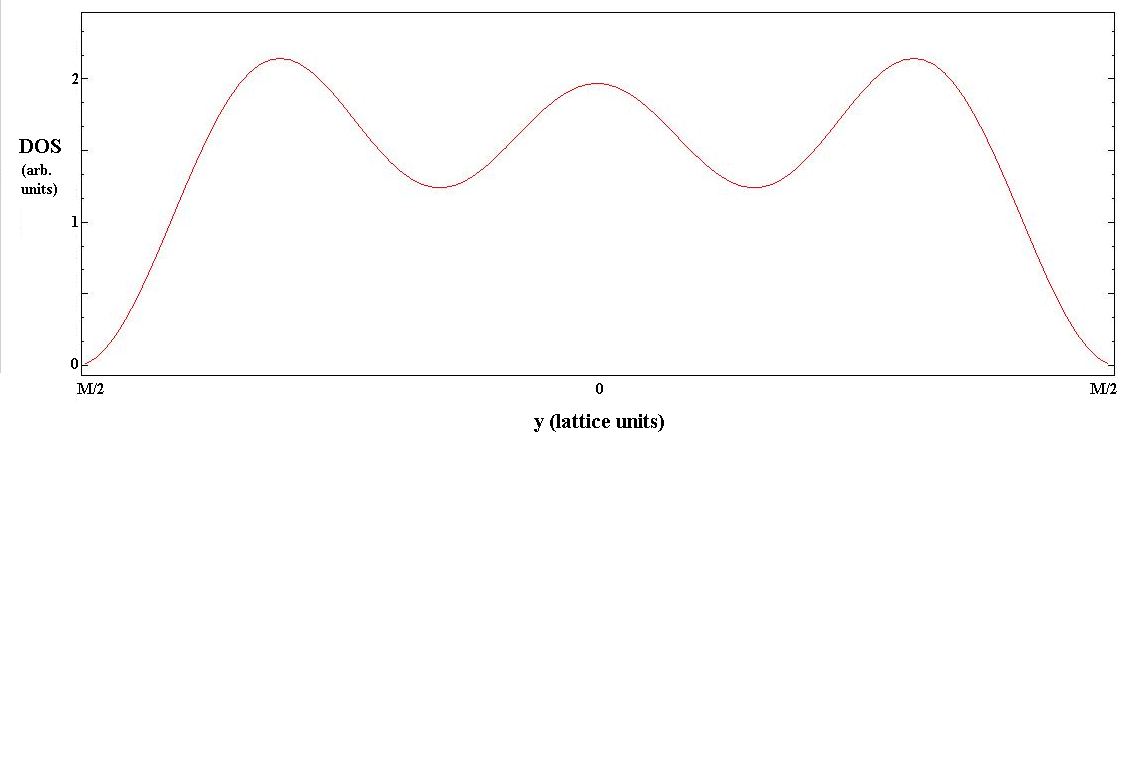}
\caption[DOS not near a subband]{\textit{Local DOS at Fermi energy well above the energy of the third subband, but below that of the fourth. Generated using numerical Green's Functions.}}\label{fig:ThreePeaksnotNS}
\end{figure}

\begin{figure}
\centering
\includegraphics[width=0.97\textwidth, viewport= 40 350 820 790]{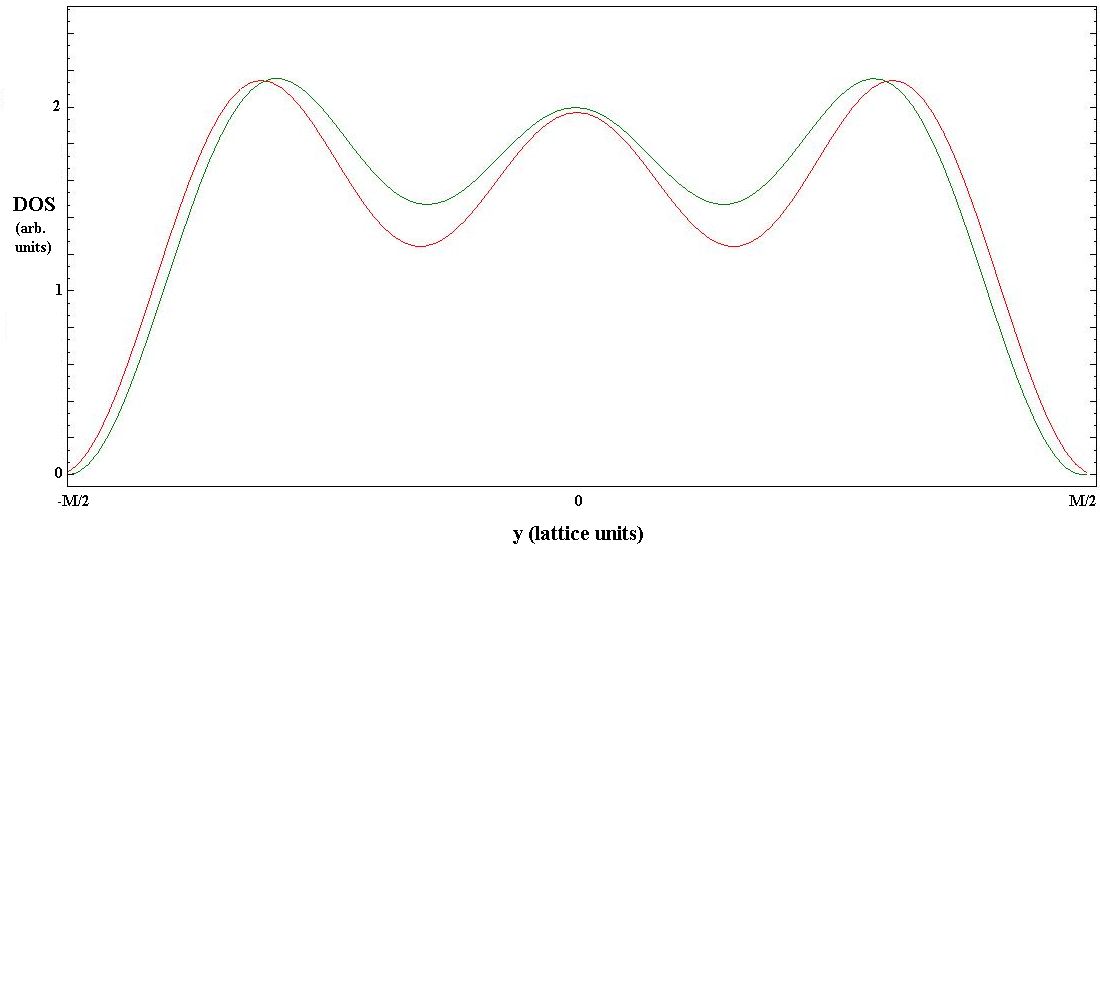}
\caption[DOS not near a subband Comparison with Analytic]{\textit{Local DOS at Fermi energy well above the energy of the third subband, but below that of the fourth, compared with plot of $\sin^2(x)+\sin^2(2x)+\sin^2(3x)$,} (green trace).}\label{fig:ThreePeaksnotNSCompare}
\end{figure}

In the opposite case, where $E$ is not near a subband's energy, one sees that no wavefunction dominates (see Figure \ref{fig:ThreePeaksnotNS}). Therefore, the
density of states slice plots should show a sum of all the subbands'
wavefunctions squared. For the case of the infinite square well 
potential, the density of states plot should look
like $\sum_{n}\sin^2(nx)$ graphs, since the solutions for a system in an infinite square well are $\sin^2(nx)$ solutions. A sum of such solutions has troughs that do not come near the x-axis like a single $\sin^2(x)$ plot; instead, both the peaks and the troughs should be well elevated from the zero-density level. As expected, using numerical Green's functions, the shape of a sum-of-sines-squared plot is readily apparent in the DOS plot (see Figure \ref{fig:ThreePeaksnotNSCompare}). 

\begin{figure}
\centering
\includegraphics[width=0.99\textwidth, viewport= 20 350 820 740]{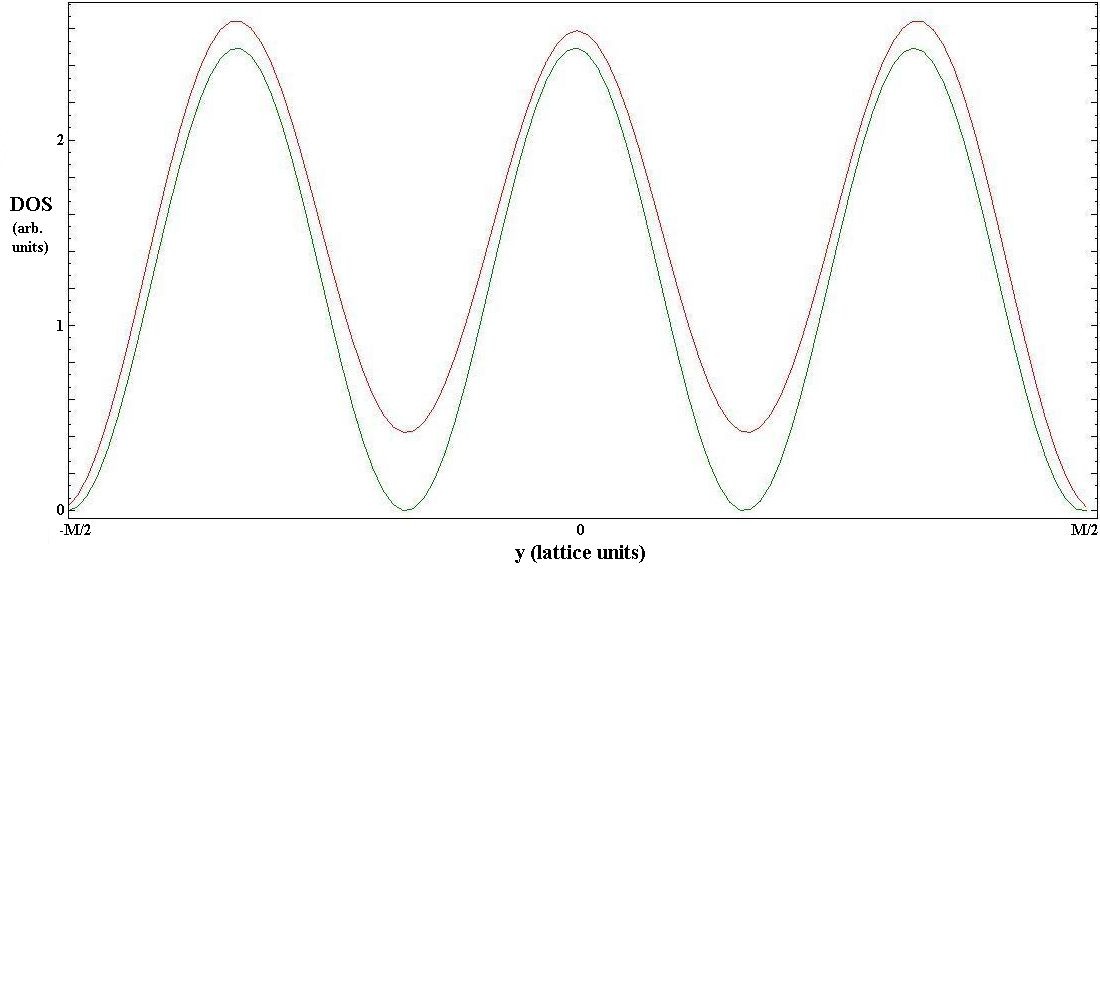}
\caption[DOS just above a subband Comparison with Analytic]{\textit{Local DOS at Fermi energy just greater than that of the third subband, compared with plot of $\sin^2(3x)$}, (green trace).}\label{fig:ThreePeaksNSCompare}
\end{figure}

In an infinite square-well potential, the results were identical for every slice of the system as can be seen from both a 2-D plot of the density of states (Figure \ref{fig:TwoDStraight}), or from taking 1-D slices from the across the system. This latter method was carried out for an $N$=50, $N$=100, $N$=200, and $N$=500 systems with ten different slices from throughout the sample---in some calculations evenly spaced and in others unevenly spaced---and each DOS was found to be identical (results are not reproduced as they simply show a single DOS since every slice's DOS overlaps perfectly). In short, for each of the density of states plots, the theoretically expected densities, peaks, and number of subbands
across all values of $E$ were generated using the numerical Green's functions method.

\subsection{Density of States at Finite Magnetic Field}

When a magnetic field is applied, the physics of the system becomes much
more interesting. What happens to the 
lowest order wavefunctions upon application of a magnetic
field has already been examined: the Lorentz force causes the wavefunctions moving in opposite direction to move against opposite walls of the system. Now the more general situation of the density of states is considered. This involves the behavior of multiple wavefunctions and the value of the Fermi energy (\ref{explicitDOS}). 

At arbitrary Fermi energy, as the magnetic field is increased, the number of subbands decreases. This results from the energy solutions to Schr\"odinger's equation under a magnetic field (\ref{SchroSubEqn}), in which $E_{0,n}\sim B$. As the increase in magnetic field causes the subband energies to grow, then at a given Fermi energy, there are fewer subbands. Thus at $B$=0, the number of subbands is at a maximum; this number gradually decreases as $B$ grows until all the subbands are depopulated. 

\begin{figure}
\centering
\includegraphics[width=0.9\textwidth, viewport= 40 420 800 760]{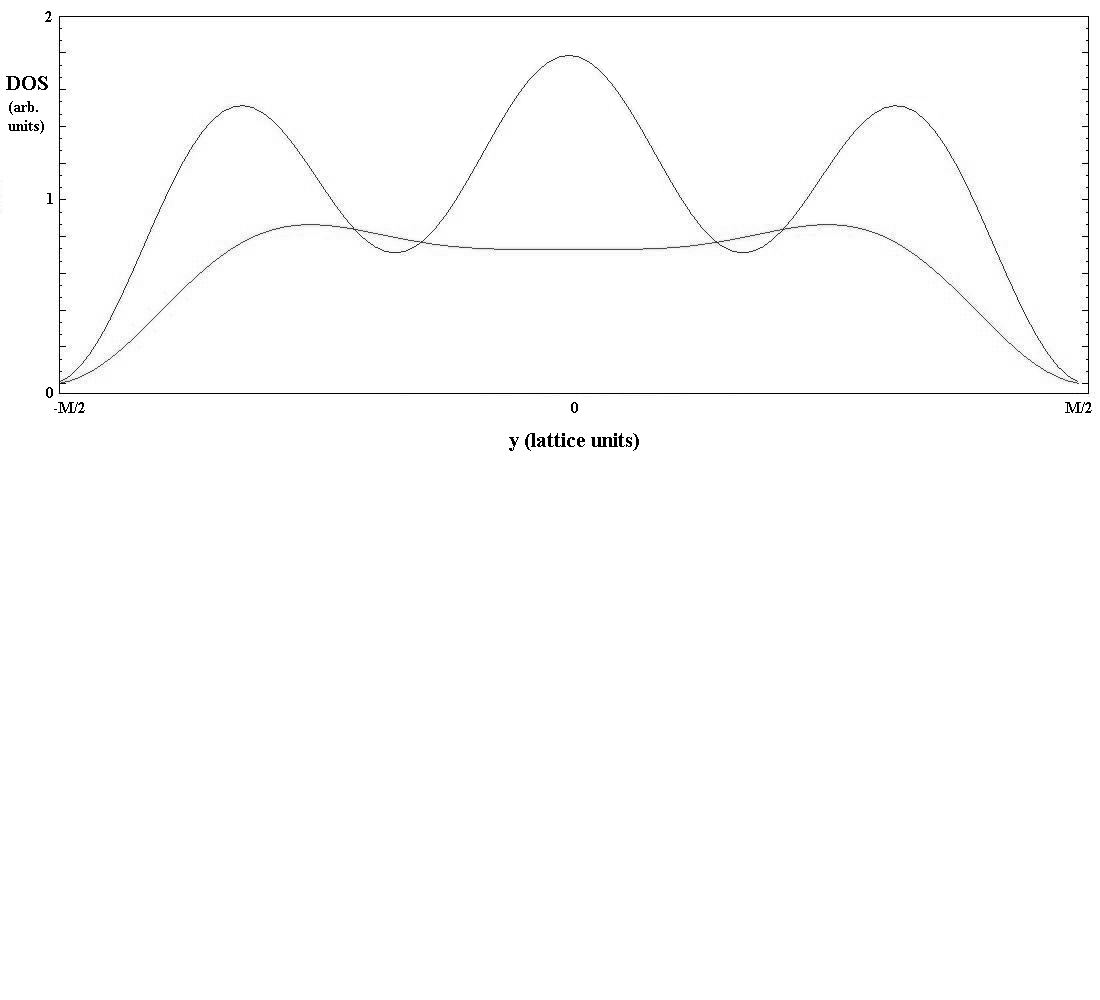}
\caption[Change in DOS with B-field]{\textit{As the magnetic field is increased, one subband depopulates at $B=0.3 T$} (lower trace), \textit{bringing the total number of subbands to two}. \textit{Increase the magnetic field further and a new hump appears, the result of the wavefunctions being pushed apart by the Lorentz force, and the second peak is the positive and negative momenta of the n=2 wavefunction overlapping; calculated at $B=1 T$} (upper trace).} \label{fig:BumpwithB}
\end{figure}

Yet, in examining the local DOS plots, one obtains results that offer greater detail than subband number alone. For while the number of subbands decreases as $B$ grows, the number of peaks in the local DOS plot increases. If one depopulates a subband at low enough B-field, there will be a drop in the number of peaks in the DOS. Yet, if one were to continue to increase the magnetic field, a large new bump would appear in the density of states calculation (see Figure \ref{fig:BumpwithB}). At first glance, this bump looks identical to a new conducting subband entering the system; however, this is not the case. For at higher magnetic-fields, beginning most noticeably around 0.5 T, the Lorentz force begins to separate \emph{all} the wavefunctions, pushing positive and negative conducting modes to opposite edges of the sample. This means that the local density of states plots contain not simply a sum of wavefunctions forming a $\sum_{n}\sin^2(nx)$ graph as before, but a sum of positive momenta wavefunctions and a sum of negative momenta wavefunctions each moving towards opposite edges of the system. As the B-field peels apart these wavefunctions, extra bumps, from new overlaps, appear in the DOS plot.

\subsection{Imaginary Band Structure}

The local DOS program can also be used to study imaginary band structure. Imaginary band structure displays interesting and not entirely understood behavior relating to the electronic properties of a quantum system and has been investigated, along with the more general complex band structure, in a variety of studies \cite{Pendry, Inkson, Yia, Blow, YQing, SMonaghan, Bross, Ghah, NStef}. Here, the imaginary band structure is derived in much the same way as the real band structure was earlier, using the eigenvalues of the transfer matrix. The imaginary bands are calculated at the $n$=0 slice. The difference from the real band calculation is that instead of using $e^{ik}=\alpha$, what is required is: 

\begin{equation}
e^{i(ik)}=\alpha \hspace{.5 cm}\longrightarrow \hspace{.5 cm} k=-\ln{\alpha}
\end{equation}

\begin{figure}
\centering
\includegraphics[width=0.32\textwidth, viewport= 80 40 380 560]{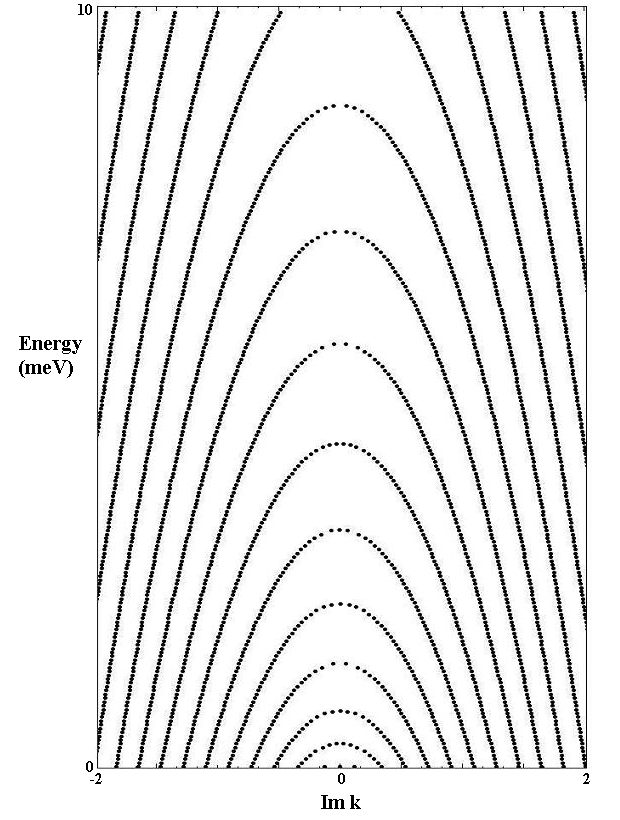}
\caption[Imaginary Band Structure at $B=0$]{\textit{Imaginary band structure, $E$ as a function of $ \textrm{Im}(k)$, at $B=0$, under the presence of no surface gates and assuming an infinite square well potential. $E$ on the vertical axis runs from 0 to 10 meV. Calculated using numerical Green's functions.}} \label{fig:CB0}
\end{figure}

\begin{figure}
\centering
\includegraphics[width=0.24\textwidth, viewport= 160 80 370 560]{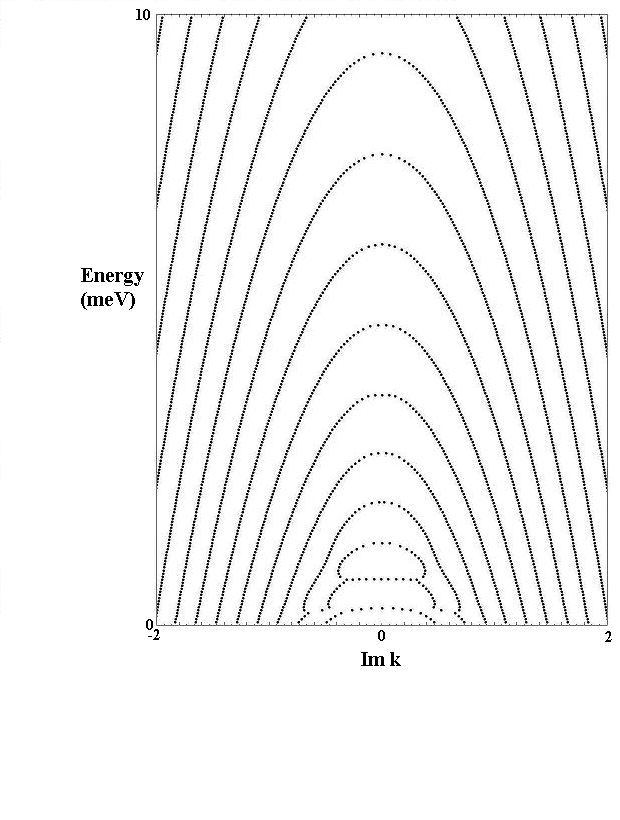}
\caption[Imaginary Band Structure at $B=0.3$]{\textit{Imaginary band structure at $B=0.3$ T, under the presence of no surface gates and assuming an infinite square well potential. Calculated using Green's functions.}} \label{fig:CB03}
\end{figure}

\begin{figure}
\centering
\includegraphics[width=0.25\textwidth, viewport= 100 30 350 600]{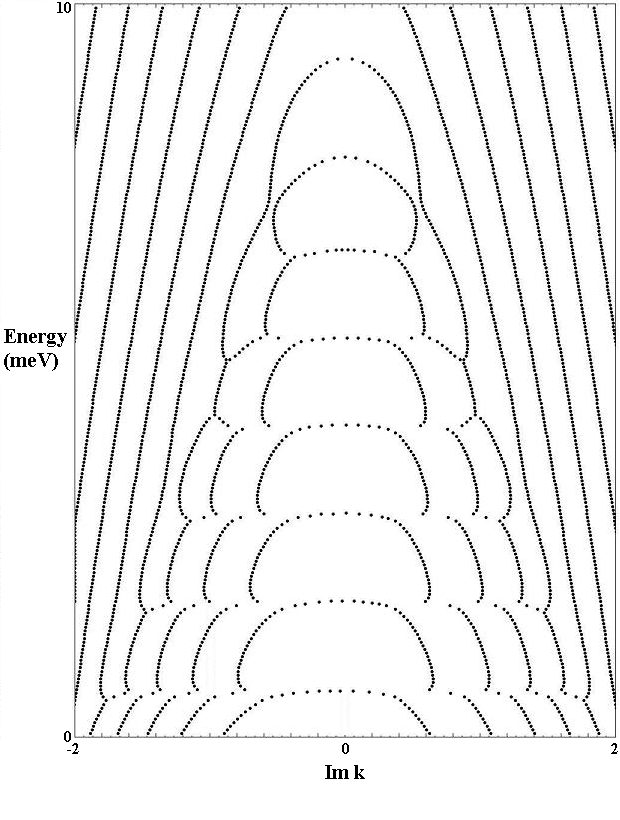}
\caption[Imaginary Band Structure at $B=0.7$]{\textit{Imaginary band structure at $B=0.7$ T, under the presence of no surface gates and assuming an infinite square well potential. Calculated using Green's functions.}} \label{fig:CB07}
\end{figure}

The imaginary band structure for a translationally-invariant potential is plotted in Figures \ref{fig:CB0}-\ref{fig:CB07}. At zero magnetic field, these negative paraboli are expected given the parabolic dispersion relation between $E$ and $k$. The maxima of these paraboli correspond precisely to the minima of the paraboli of the real band structure. As the magnetic field is turned up, however, the structure of the subbands becomes more nuanced. A ripple-like effect in the system is easily visible at just $B$= 0.3T. It is also interesting to note the asymmetries in the imaginary band structure present at just $B$=0.7 T that become even more pronounced at higher magnetic field.

\begin{figure}
\centering
\includegraphics[width=0.25\textwidth, viewport= 100 30 330 570]{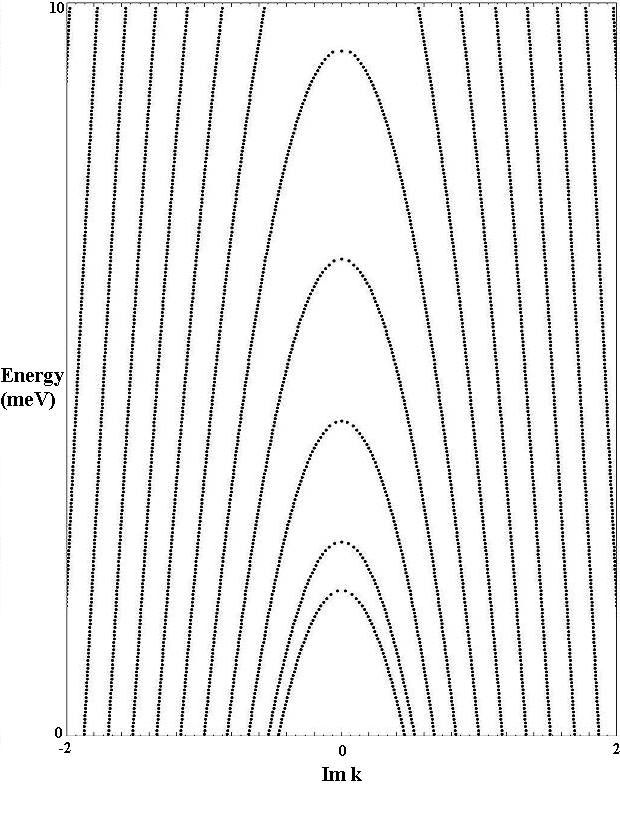}
\caption[Gated Imaginary Bands at $B=0$]{\textit{Imaginary band structure, $E$ as a function of $ \textrm{Im} (k)$, at $B=0$, under the presence of a small central surface gate. $E$ on the vertical axis runs from 0 to 10 meV. Calculated using numerical Green's functions.}} \label{fig:CBgate0}
\end{figure}

\begin{figure}
\centering
\includegraphics[width=0.25\textwidth, viewport= 100 20 330 550]{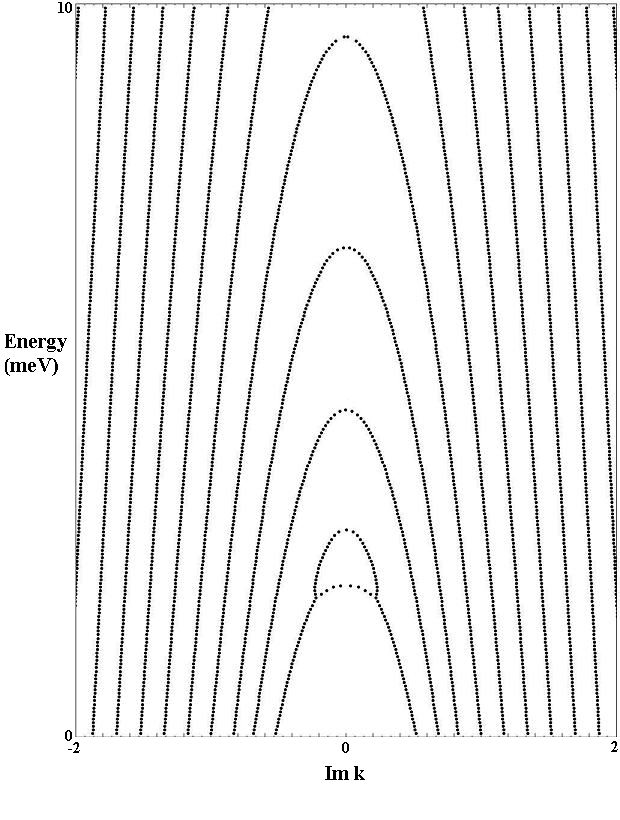}
\caption[Gated Imaginary Bands at $B=0.3$]{\textit{Imaginary band structure at $B=0.3$ T, under the presence of a small central surface gate. Calculated using numerical Green's functions.}} \label{fig:CBgate03}
\end{figure}

\begin{figure}
\centering
\includegraphics[width=0.25\textwidth, viewport= 110 40 340 580]{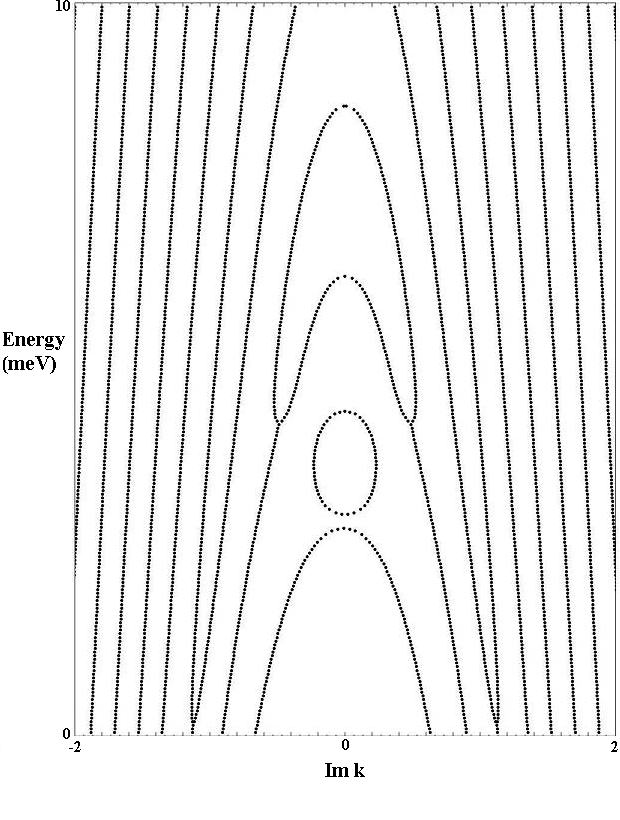}
\caption[Gated Imaginary Bands at $B=1$]{\textit{Imaginary band structure at $B=1$ T, under the presence of a small central surface gate. Calculated using numerical Green's functions.}} \label{fig:CBgate1}
\end{figure}

Especially intriguing behavior is observed when a surface gate is placed close enough to the edge of the system to impact the band structure. A similar behavior to the non-gated system for low $B$-field is observed, but the blending, i.e. the degeneracy, of the bands assumes a different behavior in this setup. A tiny rectangular gate of length 6.5 nm and width 26 nm, was centered in the middle of a 1-D system of length 65 nm and width 131.3 nm. The resulting imaginary band behavior can be seen in Figures \ref{fig:CBgate0}-\ref{fig:CBgate1}. 

Initially, the nature of the points where two imaginary subbands merge was investigated, in particular where subbands 3 and 4 or 5 and 6 overlap. The local density of states above and below these points of degeneracy was plotted, and each 2-D density map was subtracted from the other, but no pattern emerged. The density differences between a plot for a Fermi energy above the degeneracy and for a plot below the degeneracy could be just as prominent as the density differences between two plots both energetically above the degeneracy point or both energetically below it. 

\begin{figure}
\centering
\includegraphics[width=0.3\textwidth, viewport= 150 40 340 620]{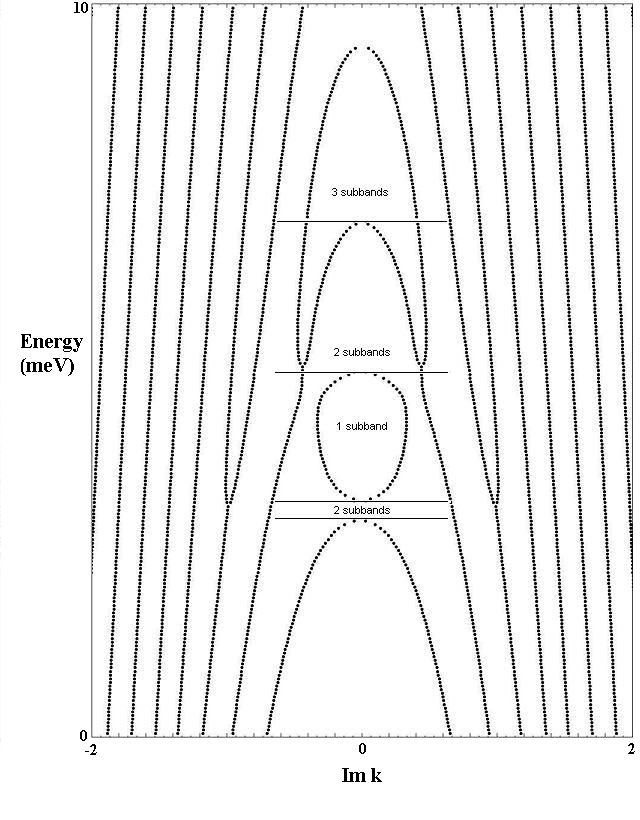}
\caption[Gated Imaginary Bands at $B=1.2$]{\textit{Imaginary band structure at $B=1.2$ T, under the presence of a small central surface gate. The horizontal lines demarcate the Fermi energies as whose values a subband is added (or, in the strange case, subtracted). Calculated using numerical Green's functions.}} \label{fig:CBgate12}
\end{figure}

What turned out to be the most interesting facet of the imaginary band structure was that at high magnetic field the number of conducting modes depended upon the bottom portion of the second imaginary subband. Since the number of conducting modes depends only upon the real number of subbands this seemed a strange property. Between the top of the first imaginary subband and the bottom of the second subband the system had two conducting modes (see Figure \ref{fig:CBgate12}). Above the bottom of the second subband and below the top of the second subband, the system had just one conducting mode, then returned to two again at the top of second imaginary subband.

\begin{figure}
\centering
\includegraphics[width=0.3\textwidth, viewport= 170 70 340 590]{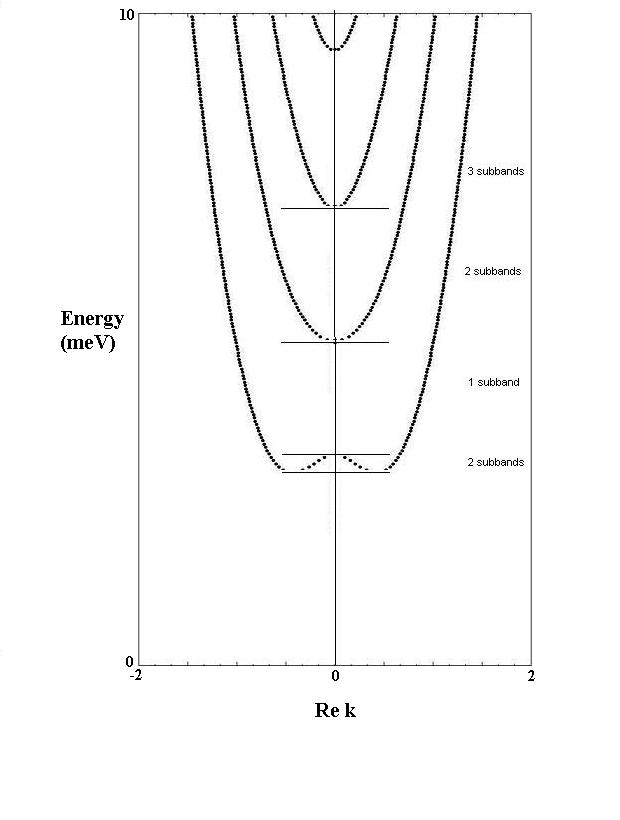}
\caption[Gated Real Bands at $B=1.2$]{\textit{Real band structure at $B=1.2$ T, under the presence of a small central surface gate. The horizontal lines demarcate the Fermi energies as whose values a subband is added (or, in the strange case, subtracted). Calculated using numerical Green's functions.}} \label{fig:RBgate12}
\end{figure}

A plot of the real subbands under the influence of the same nearby gate revealed the reason for this behavior (see Figure \ref{fig:RBgate12}). The gate distorts the first real subband, causing the energy values for small but non-zero $k$ to be less than the values of the subband at $k$=0. As a result, selecting an energy level below the hump at $k$=0, but above the two lowest minima on either side of $k$=0, yields four intersections, which corresponds to two subbands (each with a positive and negative momentum solution). When the energy value just above $k$=0 is reached, there are only two intersections and hence, only one subband. This continues until the second subband is reached at which point there are again two conducting modes. 

In this way, the imaginary band structure reflects the real band structure. The closed-off ellipse of the second imaginary subband reflects the symmetric dips at small $k$ on the real subbands. The reason for these dips in the real subband structure is likely the effect of the gate: states with slightly positive or negative momenta will be pushed to either side of the gate as $B$ is increased. Those states with no momentum ($k$=0), however, will flow directly into the gate even at high $B$. Thus, the gate causes real subbands to appear for small momenta at lower energies than they do for zero momenta, resulting in the anomalous gated imaginary band behavior.

\section{The Transmission Coefficients Program}

The transmission coefficients calculation program shared a great deal in common with the local density of states program; indeed, almost all of the initial work is identical. The aim of the program is to calculate the transmission, $T_{ij}$, and reflection, $R_{ii}$, coefficients, for leads $i$ and $j$ of the Landauer-B\"{u}ttiker formalism. By calculating these values the degree of transport through the quantum sample---a crucial characteristic of 1D dynamics---is determined.

Like the local DOS program, the transmission program begins with the selection of inputs: $M$, $N$, $a$, $E$, and $B$. The program is easily adjusted to calculate the transmission coefficients as a function of magnetic field as is done in Chapter Five. The effective mass used was again that of GaAs, $0.067e$. The constants $\nu$, $\epsilon$, and $is$ are used as before, as are $\mathbf{V}$, $\mathbf{V}^\dag$, and $\mathbf{Z}$. The Hamiltonian and its dependence upon an external effective potential was again implemented. Eigenvalues and eigenvectors of the transfer matrix were found and sorted, and $\mathbf{G}^{-\infty}_{0,0}$ was calculated. 

Here the process diverges from that used for calculating the local density of states. There, since the local DOS at every lattice point was sought, one had to iterate all the way across the system in order to find the DOS for a single slice and therefore there were $N$ total iterations across the system. By contrast, finding the transmission coefficients requires iterating across the system only once. 

In the local DOS program, the end goal is to find $\mathbf{G}^{(N)}_{w,w}$ matrix; in the transmission coefficients program, the goal is to find matrices $\mathbf{G}^{(N)}_{N,N}$ (of (\ref{FiveA})) and $\mathbf{G}^{(N)}_{0,N}$ (of (\ref{FiveC})), then add on the leads. The latter of these matrices is particularly important as it relates the Green's function at one end of the system to the other, i.e. transmission. To determine them, $\mathbf{G}^{-\infty}_{0,0}$ is inserted into (\ref{FiveA}) as before to find $\mathbf{G}^{(1)}_{1,1}$, which, in combination with $\mathbf{G}^{-\infty}_{0,0}$ yields $\mathbf{G}^{(1)}_{0,1}$. The repetition of this process $N$ iterations later gives $\mathbf{G}^{(N)}_{N,N}$ (of (\ref{FiveA})) and $\mathbf{G}^{(N)}_{0,N}$.

This iterative process is followed by a series of matrix multiplications that are carried out in order to string together Green's functions that fully describe transmission across the system. The many matrix products are then incorporated into an expression that determines the transmission, $T_{ij}$ from any one subband $i$ on one side of the system to subband $j$, on the other. The expression is given by the sum of four terms (with Green's functions condensed wherever possible):

\begin{equation}\label{TransCodeEq}
\begin{array}{cccc}
&1)& \mathbf{U}^{+*}\mathbf{G}_{0,N}^{+\infty}\mathbf{V}\mathbf{G}_{N+1,N+1}^{+\infty}\mathbf{U}^{+}\alpha^{+}\mathbf{V}\mathbf{V} \\
&2)&\mathbf{U}^{+*}\alpha^{+*}\mathbf{G}_{0,0}^{-\infty}\mathbf{V}\mathbf{G}_{0,N}^{+\infty}\mathbf{V}\mathbf{G}_{N+1,N+1}^{+\infty}\mathbf{V}^{\dag}\mathbf{G}_{N,N+1}^{+\infty}\mathbf{U}^{+}\mathbf{V}^{\dag}\mathbf{V}^{\dag} \\
&3)&-\mathbf{U}^{+*}\alpha^{+*}\mathbf{G}_{0,0}^{-\infty}\mathbf{V}\mathbf{G}_{0,N}^{+\infty}\mathbf{V}\mathbf{G}_{N+1,N+1}^{+\infty}\mathbf{U}^{+}\alpha^{+}\mathbf{V}^{\dag}\mathbf{V} \\
&4)&-\mathbf{U}^{+*}\mathbf{G}_{0,N}^{+\infty}\mathbf{V}\mathbf{G}_{N+1,N+1}^{+\infty}\mathbf{V}^{\dag}\mathbf{G}_{N,N+1}^{+\infty}\mathbf{U}^{+}\mathbf{V}\mathbf{V}^{\dag}
\end{array}
\end{equation}

This expression is derived from the work of Appendix B of \cite{Baranger}, where the lattice form of the transmission coefficients in terms of Green's function is derived. An operator related to the current density, $K_{op}(n)$, is introduced: 

\begin{equation}\label{BSAppTwo}
K_{op}(n)\equiv\frac{e}{i\hbar}\sum_m\left(V_{n,m} \vert n,m \rangle \langle n+1, m \vert - V^\dag_{n,m} \vert n+1, m\rangle \langle n, m \vert \right)
\end{equation}
Where $V_{n.m}$ and $V^\dag_{n,m}$ are the $m$th entry of matrices $\mathbf{V}$ and $\mathbf{V^\dag}$, respectively, calculated at slice $n$; $\vert n+1,m \rangle$ is an eigenvector one lattice point to the right of the eigenvector at point $n,m$. $K_{op}(n)$ can be thought of as the current that passes between slice $n$ and slice $n+1$. Furthermore, $K_{op}(n)$ is very similar to the current density operator $J_{op}(n)$ used to link the Green's function formalism to the scattering formalism (see \cite{Baranger} for details), and therefore can be related to the conductance coefficients, $g_{ij}$, between leads $i$ and $j$. This in turn leads to an expression for the transmission coefficients, where $\mathbf{G}_{i,j}$ is the Green's function connecting leads $i$ and $j$ across the entire system:

\begin{equation}\label{BSAppOne}
t_{ij,ab}=-\frac{i\hbar}{e^2} \langle \psi_a^+ \vert K_{op}(n')\mathbf{G}_{i,j}K_{op}(n) \vert \psi_b^- \rangle \hspace{.8 cm} \textrm{ where } i \neq j
\end{equation} 
Here, $\psi_a^+$ and $\psi_b^-$ are the eigenvectors of subbands $a$ and $b$ moving in the positive and negative directions, respectively. Inserting (\ref{BSAppTwo}) into (\ref{BSAppOne}), one obtains four terms, which when expanded give the terms of (\ref{TransCodeEq}). The $\psi_a$ and $\psi_b$ vectors become the $\mathbf{U_+}$ and $\mathbf{U_-}$ entries; the $V_{n,m}$ terms correspond to $\mathbf{V}$ and $\mathbf{V^\dag}$; the eigenstate and Green's function products form the basis of the Green's functions multiplications in (\ref{TransCodeEq}). This is the source of the transmission program.

Once (\ref{TransCodeEq}) is applied, then for the case of only two leads, one obtains a transmission matrix whose entries are simply $t_{ij}$. Then, as given by the Landauer formula (\ref{LanCorrect}), one sums over every value of the matrix. This process calculates the total conductance, across all subbands, through the system.

A similar method is applied to find the reflection coefficients. The relevant expression for calculating the reflection coefficients from lead $i$ back into lead $i$ is $R_{ii}$. It is given by the sum of four terms (with Green's functions condensed wherever possible):

\begin{equation}\label{ReflCodeEq}
\begin{array}{cccc}
&1)& \mathbf{U}^{-*}\mathbf{G}_{N+1,N+1}^{+\infty}\mathbf{U}^{+}\alpha^{+}\mathbf{V}\mathbf{V} \\
&2)&\mathbf{U}^{-*}\alpha^{-*}\mathbf{G}_{N,N}^{+\infty}\mathbf{V}\mathbf{G}_{N+1,N+1}^{+\infty}\mathbf{V}^{\dag}\mathbf{G}_{N,N+1}^{+\infty}\mathbf{U}^{+}\mathbf{V}^{\dag}\mathbf{V}^{\dag} \\
&3)&-\mathbf{U}^{-*}\alpha^{-*}\mathbf{G}_{N,N}^{+\infty}\mathbf{V}\mathbf{G}_{N+1,N+1}^{+\infty}\mathbf{U}^{+}\alpha^{+}\mathbf{V}^{\dag}\mathbf{V}\\
&4)&-\mathbf{U}^{-*}\mathbf{G}_{N+1,N+1}^{+\infty}\mathbf{V}^{\dag}\mathbf{G}_{N,N+1}^{+\infty}\mathbf{U}^{+}\mathbf{V}\mathbf{V}^{\dag}
\end{array}
\end{equation}

This expression is derived from \cite{Baranger}, where the reflection coefficients equivalent to (\ref{BSAppOne}) can be found. The result (\ref{ReflCodeEq}) can be derived using identifications similar to those listed above for the transmission coefficients. As a check on the system, the sum of the total transmission and reflection coefficients should be equal to the number of total conducting subbands. This was found to be true.

Results from the transmission program for an ordinary system matched theoretical expectations. For the case of a translationally-invariant potential such as in Fig. \ref{fig:TwoDStraight}, the transmission coefficients program produces three transmitted and zero reflected subbands, as expected. Also as predicted, as the magnetic field is increased leading to depopulation, the program calculates that the number of transmitted subbands falls to zero while the number of reflected subbands stays fixed at zero. Under the presence of a non-translationally invariant potential---for example, a surface gate in the middle of the channel---some subbands are reflected. The reflection coefficient will actually decrease at higher magnetic field, both because of magnetic deopopulation and because the Lorentz force pushes the subbands against the the sample's edges, allowing them to move around the central gate and conduct to the opposite side of the sample.

What is intriguing about the transmission coefficients program, however, stems not from its values in various geometries alone, but from comparing its results with those from the density of states program. Here, the limits of experimental measurement---which can only determine conductance---are made evident. The density of states model can provide details of quantum systems unseen by direct experiment and previously understood only by inference. The next and final chapter considers two of the many future directions to which these two programs could be put in combination to probe frontiers in low-dimensional quantum physics.

\chapter{Future Directions}
\section{Introduction: Integrating the DOS and Transmission Coefficients Programs}

As the local DOS is a fundamental property of a quantum system and the conductance through a 1D system is the fundamental limit of studying electronic transmission, there are innumerable future investigations in which the programs presented in the previous chapter could prove of service. Any experiment relating to transport through a narrow constriction (narrow, because the wider the system is, the exponentially longer the calculations become), inclusive of most geometries, attempting to probe conductance, thermopower, magnetic-field effects, energy eigenstate distributions, and other quantum properties could be modeled with these two programs. Below, it is examined how the models presented in this thesis could be applied to two current realms of inquiry: antidot behavior and zero-dimensional to one-dimensional tunneling.

\section{AntiDot Behavior}

\begin{figure}
\centering
\includegraphics[width=0.5\textwidth, viewport= 0 0 250 250]{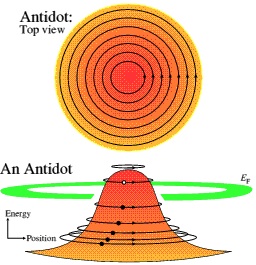}
\caption[An Antidot]{\textit{An Antidot is a potential hump frequently placed in the middle of a split-gate 1D channel. Circular orbits represent edge states moving around the antidot. Image from \cite{website}.}} \label{fig:Antidot}
\end{figure}

An antidot is a bump in the effective potential in a 1D channel, producing a region from which electrons are excluded (see Figure \ref{fig:Antidot}). It can be formed by placing a very small gate in the middle of a split-gate device. Antidots contain many interesting properties and are ripe for the study of magnetoconductance, scattering, and tunneling behaviors \cite{Sachr, Kirc, Kircz}. They have been studied in a variety of situations, often with intriguing results concerning Quantum Hall edge states as well as spin properties \cite{KataThree, KataOne, KataTwo}. 

Antidot systems are excellent candidates for analysis by the local DOS and transmission coefficients programs and might be especially well deployed in a single antidot system described in \cite{MaceBarnes}. Here, an eigenstate moves around the antidot. Tunneling is present both across the channel, leading to increased transmission, and tranverse to the channel, leading to increased reflection. Varying an applied magnetic field causes resonant dips and peaks in the transmission and reflection coefficient values. The voltage on the antidot (and therefore the radius of the antidot) is an essential parameter of the system, because changes in the antidot's size affect the eigenstates and therefore the degree of tunneling that is possible.

\begin{figure}
\centering
\includegraphics[width=0.6\textwidth, viewport= 200 40 720 420]{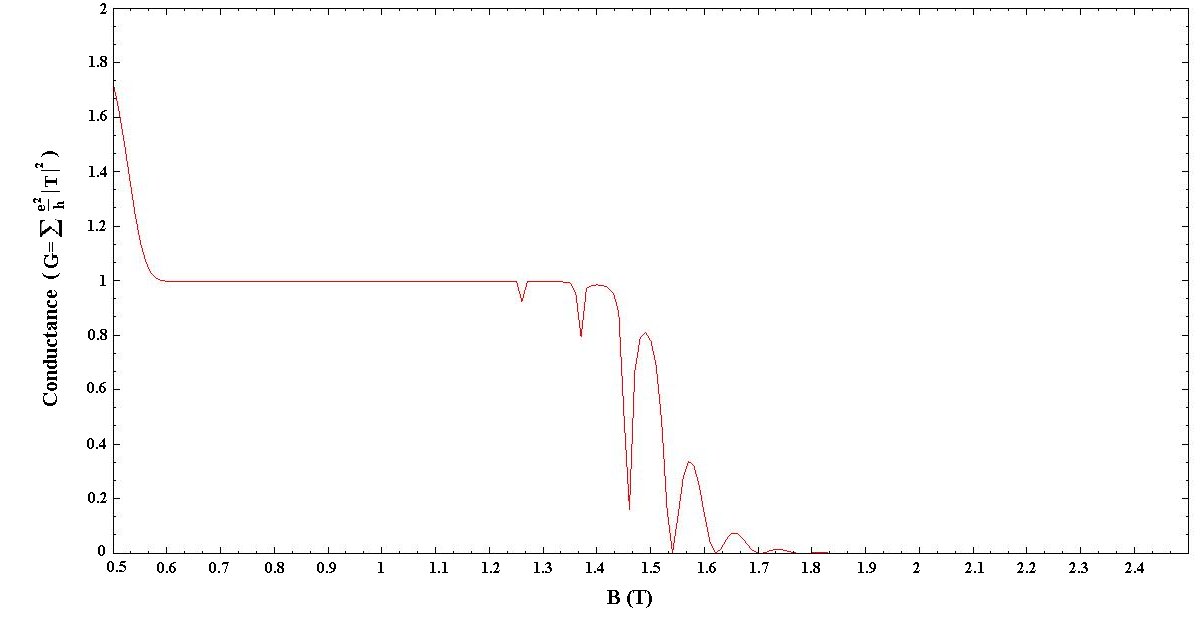}
\caption[Conductance Through an Antidot System]{\textit{Conductance as a function of magnetic field through an antidot system. A value of one means one subband (one moving left, one moving right) passing through the system. Voltage on antidot=-3.5 mV. Note the resonant dips due to increased reflection.}} \label{transdipsSmallV}
\end{figure}

\begin{figure}
\centering
\includegraphics[width=0.6\textwidth, viewport= 200 40 720 420]{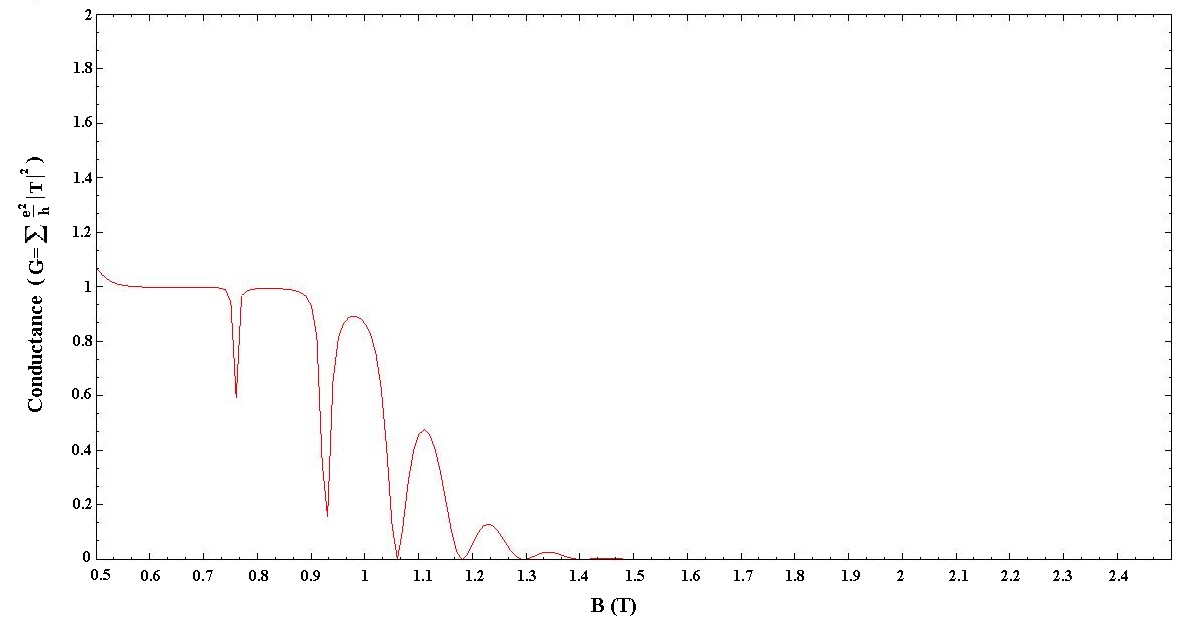}
\caption[Conductance Through an Antidot System II]{\textit{Conductance as a function of magnetic field through an antidot system. Voltage on antidot=-4.5 mV. Note how the more negative voltage lowers the magnetic field at which resonant reflection begins to occur because there is greater subband overlap at lower B-field than in Figure \ref{transdipsSmallV}.}} \label{transdipsBigV}
\end{figure}

Applying the transmission and local DOS programs to calculate the transmission coefficients as a function of magnetic field revealed sharp resonant dips at places where tunneling transverse to the channel suddenly peaked. These peaks cause transmission across the channel to decrease (see Figure \ref{transdipsSmallV}). One would expect making the voltage on the central gate more negative (thereby increasing the size of the antidot), would push the eigenstates around the dot closer to the ones at the channel's edges. This increased proximity would increase vertical tunneling and therefore cause an onset of resonant vertical tunneling at lower magnetic field. This is precisely what was found (see Figure \ref{transdipsBigV}, where the resonant dips are apparent at lower B-field).

\begin{figure}
\centering
\includegraphics[width=0.57\textwidth, viewport= 140 0 600 400]{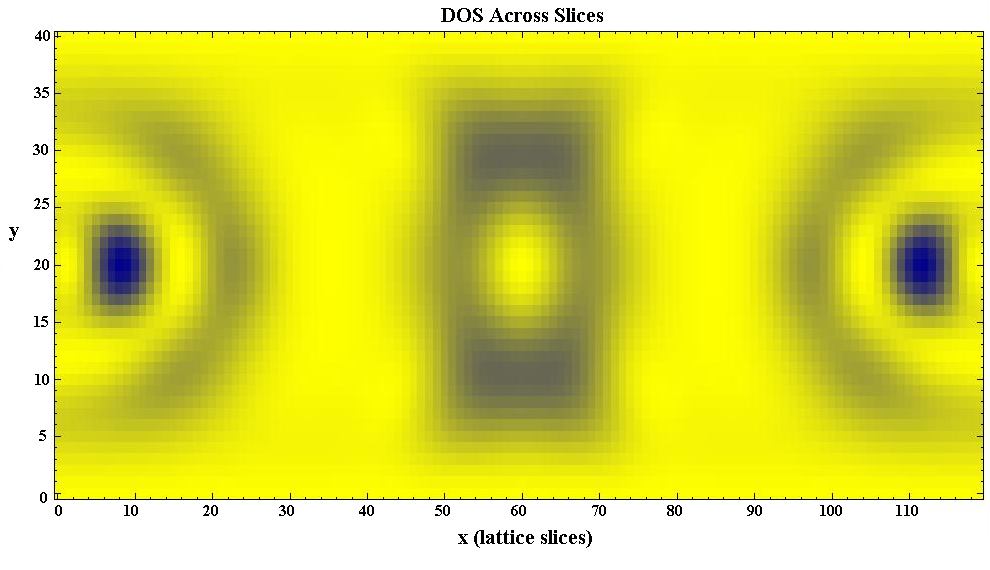}
\caption[Transmission Dip DOS Plot]{\textit{Local DOS plot for $B=.93$ T. Blue represents high local DOS, yellow low local DOS. Transmission is imperceptible here.}} \label{fig:NoTransmit}
\end{figure}

\begin{figure}
\centering
\includegraphics[width=0.57\textwidth, viewport= 140 0 600 400]{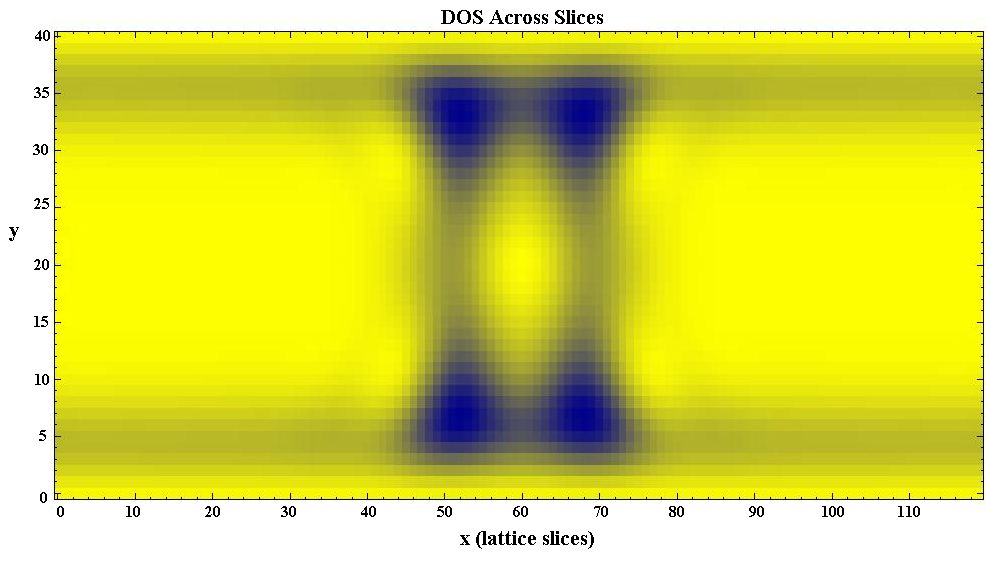}
\caption[Transmission Full DOS Plot]{\textit{Local DOS plot for $B=.96$ T. Transmission is evident here in the blue lines that cross the entire system which are the subbands pushed to the edges by the Lorentz force.}} \label{fig:Transmit}
\end{figure}

Results of greater interest emerge when the local density of states program is applied under the same conditions to the same system. This provides for a portrait of the sample's transmissive behavior. The decrease in tunneling across the system is evident in the vanishing cross-channel density of states at the values of the resonant dips (for example, see Figure \ref{fig:NoTransmit}, the local DOS at $B=.93$ T). The edge states at these resonances are not transmitted across the channel. If one looks at a plot of the local DOS where the conductance is nearly one again, however, the local DOS makes plain that transmission across the channel is once again present (see Figure \ref{fig:Transmit}). 

The local DOS was calculated at dozens of magnetic fields in the system. For a $N$=200 system of width 250 nm, calculating each local DOS plot for a given magnetic field takes under five minutes on a standard desktop computer, and the expediency of the code itself could always be improved. Applying a whole host of new conditions could be simply done and studies of edge states, tunneling, spin properties, and other phenomena could be carried out and analyzed rapidly by looking at the transitions of the transmission coefficients and local DOS across a range of energetic or magnetic perturbations.

\section{0D-1D Tunneling}
A second active area of research where this pair of programs could be deployed is the investigation and modeling of zero-dimensional to one-dimen-sional tunneling systems. Most recently these systems have been probed using a surface-acoustic wave (SAW) to create a \emph{moving} quantum dot---a zero-dimensional, fully confined electronic system---that then tunnels into a neighboring 1DEG \cite{Astle}. These tunneling oscillations could prove crucial to creating a qubit, the backbone of a quantum computation device.

\begin{figure}
\centering
\includegraphics[width=0.79\textwidth, viewport= 80 280 750 740]{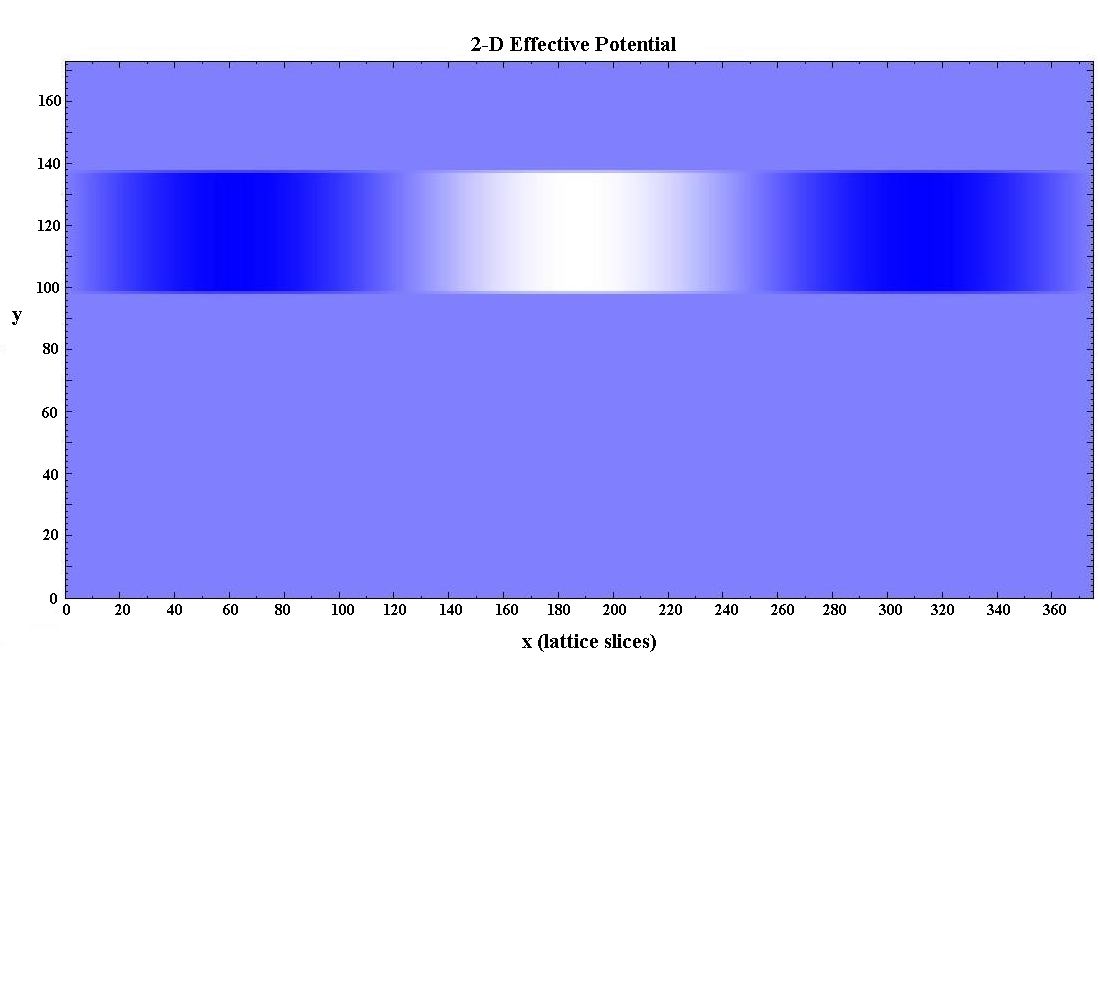}
\caption[Effective Potential for 0D-1D tunneling]{\textit{Effective Potential for SAW in 0D-1D tunneling system, using (\ref{PSAW}). White demarcates low effective potential, light blue medium potential, dark blue high effective potential. The SAW is clearly confined to the upper channel with the moving quantum dot at the upper channel's center.}} \label{fig:plotforSawPotential}
\end{figure}

To model such a device, one defines a series of surface gates using the formalism derived from \cite{DaviesL}. In essence, one creates two 1D quantum channels, side-by-side, with a small, electrostatically-defined potential hump separating them; this hump is controlled by a gate, the gate voltage in turn controlling the degree of tunneling between channels. In one of the 1D channels, the potential due to a surface acoustic wave is modeled as a sine wave moving in the x-direction and boundaries in the y-direction decaying sharply like Fermi-Dirac functions against the edges of the 1D channel. In initial calculations performed, the potential was approximated as (see Figure \ref{fig:plotforSawPotential}):

\begin{equation}\label{PSAW}
P_{Saw}= \frac{A\sin(\frac{2\pi(x-x_{mid})}{\lambda/4}-\frac{\pi}{2})}{(e^{10(y-y_{top})}+1)(e^{-10(y-y_{bottom})}+1)}
\end{equation}
where $y_{bottom}$ and $y_{top}$ are the electrostatic boundaries of the 1DEG channel in which the SAW is present, $x_{mid}$ is the middle slice of the system, $A$ is the amplitude of the saw in mV, and $\lambda$ is the wavelength of the SAW, on the order of 1 $\mu$m. Using (\ref{PSAW}) as the effective potential, adding in the effect of the gates, and then choosing suitable lattice dimensions for the channel, the numerical models presented above tap into stores of information about the expected theoretical behavior of the resonant tunneling system. For example, one can analyze the distribution of the local density of states, identifying potentially anomalous behaviors in wavefunction distribution. New gate geometries can be quickly modeled and potential problems in transmission immediately identified and calculated.

The effective potential of (\ref{PSAW}) considers the SAW at a single snapshot in time, at the moment when it has a potential minimum midway through the channel. A time-dependent potential could be incorporated into the Green's function model to make it an even more powerful probe of the system. In this way, the truly dynamic nature of the quantum dot due to the SAW could be modeled. One simple method of probing the local DOS behavior as the SAW moves through the system would be to run the calculations for the SAW with its potential minimum at several hundred different lattice points, creating a ``moving" picture of the dynamics of this tunneling system.

\section{Conclusions}

This thesis has presented a model for calculating the local density of states and transmission coefficients. The foundation of the program---the iteration of numerical Green's functions to obtain relevant energetic information about the system---has been derived and explicated. The basics of semiconductor transport in a 1DEG, employing the Landauer-B\"{u}ttiker formalism, have also been elucidated. 

Two numerical programs have been proposed and presented. The extremely high degree of accuracy of these programs has been demonstrated in several manners. The programs' results, most notably the local DOS program, match analytic results for local density of states, real band structure, and waveband nature and behavior with great precision. The local density of states program also evinces expected theoretical behavior in terms of number of subbands, peak positions, and depopulation under the influence of a magnetic field and free of its influence. Furthermore, it has been shown that the local DOS program can be used to bring to light interesting difficulties in the imaginary band structure and leaves room for others to be investigated. The local DOS can be used to quickly provide detailed schematics of wavefunction behavior under the presence of almost any arbitrary  effective potential. Together, the transmission program and the local DOS program are easily adaptable to the investigation of numerous ongoing experimental inquiries, including 0D-1D tunneling and antidot systems.

\singlespacing

\bibliography{Total6}

\doublespacing
\appendix
\chapter{Derivation of Iterative Equations}

The four iterative Green's functions equations for relating two slices of the quantum systems are given by:

\begin{eqnarray}
\mathbf{G}^{(n+1)}_{n+1,n+1}&=&[\mathbf{Z}-\mathbf{H}_{n+1}-\mathbf{V^\dag} \mathbf{G}^{(n)}_{n,n}\mathbf{V}]^{-1} \label{FiveAA}\\
\mathbf{G}^{(n+1)}_{i,j}&=&\mathbf{G}^{(n)}_{i,j}+\mathbf{G}^{(n)}_{i,n} 
\mathbf{V}\mathbf{G}^{(n+1)}_{n+1,n+1}\mathbf{V^\dag}\mathbf{G}^{(n)}_{n,j} \hspace{.8 cm} (i, j\leq N) \label{FiveBA}\\
\mathbf{G}^{(n+1)}_{i,n+1}&=&\mathbf{G}^{(n)}_{i,n}\mathbf{V} \mathbf{G}^{(n+1)}_{n+1,n+1}\hspace{.8 cm}(i\leq N) \label{FiveCA}\\
\mathbf{G}^{(n+1)}_{n+1,j}&=&\mathbf{G}^{(n+1)}_{n+1,n+1}\mathbf{V^\dag} \mathbf{G}^{(n)}_{n,j} \hspace{.8 cm} (j\leq N)\label{FiveDA}
\end{eqnarray}

Equation (\ref{FiveCA}) is derived directly from the Dyson equation for the case $j=n+1$. The first term drops out because the $n$th Green's function does not contain an entry for $\textbf{G}^{(n)}_{i,n+1}$. This is because by the $n$th iteration, there is not yet an $n+1$ column. Only once $\textbf{G}^{(n+1)}$ is calculated can (\ref{FiveCA}) provide the values of the last column of the total Green's function matrix---save the very last row's entry.

This final row, final entry of the Green's function given by (\ref{FiveAA}) is derived from (\ref{FiveCA}). First $i,j=n+1$ are plugged into the Dyson equation. Second, the result of (\ref{FiveCA}) is substituted in for $\mathbf{G}^{(n+1)}_{n,n+1}$ yielding:

\begin{equation}\label{FiveAStep}
\mathbf{G}^{(n+1)}_{n+1,n+1}=\mathbf{G}^{(n)}_{n+1,n+1} + \mathbf{G}^{(n)}_{n+1,n+1}\mathbf{V}^\dag \mathbf{G}^{(n)}_{n,n}\mathbf{V}\mathbf{G}^{(n+1)}_{n+1,n+1}
\end{equation}
where $\mathbf{V}^\dag$ is used instead of $\mathbf{V}$ because it connects $ \mathbf{G}^{(n)}_{n+1,n+1}$ to a lattice slice to the left, $\mathbf{G}^{(n)}_{n,n}$. Subtracting the right-most term from (\ref{FiveAStep}), factoring out $\mathbf{G}^{(n+1)}_{n+1,n+1}$ on the left-side and multiplying by $(\mathbf{G}^{(n)}_{n+1,n+1})^{-1}$ one arrives at:

\begin{equation}\label{FiveAStepTwo}
\mathbf{G}^{(n+1)}_{n+1,n+1} = ((\mathbf{G}^{(n)}_{n+1,n+1})^{-1} - \mathbf{V}^\dag\mathbf{G}^{(n)}_{n,n}\mathbf{V})^{-1}
\end{equation}
which, by the definition of the Green's function reduces to (\ref{FiveAA}).

Like (\ref{FiveCA}), (\ref{FiveDA}) can be derived straightforwardly from the Dyson formula. If $i=n+1$, then as a result of the ($j \leq N$) constraint of (\ref{FiveDA}), the relevant Dyson formulation will be: $ \mathbf{G}^{(n+1)}_{n+1,j}=\mathbf{G}^{(n)}_{n+1,j}+ \mathbf{G}^{(n)}_{n+1,n+1}\mathbf{V}^\dag \mathbf{G}^{(n+1)}_{n,j}$. As with the derivation of (\ref{FiveCA}), and for the same reason, the first term does not exist. Using (\ref{DysonM}) to substitute for $\mathbf{G}^{(n+1)}_{n,j}$ and multiplying by $(\mathbf{G}^{n}_{n+1,n+1})^{-1}$ the result is obtained:

\begin{equation}\label{FiveDStep}
((\mathbf{G}^{(n)}_{n+1,n+1})^{-1}-\mathbf{V}^\dag \mathbf{G}^{(n)}_{n,n}\mathbf{V})\mathbf{G}^{(n+1)}_{n+1,j}=\mathbf{V}^\dag \mathbf{G}^{(n)}_{n,j}
\end{equation}

The parenthetical term on the left-hand side is $(\mathbf{G}^{(n+1)}_{n+1,n+1})^{-1}$ according to (\ref{FiveAStepTwo}). Inputting this Green's function into (\ref{FiveDStep}) and multiplying both sides by its inverse, returns (\ref{FiveDA}). 

The final equation, (\ref{FiveBA}), is derived from (\ref{FiveDA}) in a single step. Employing (\ref{DysonM}) directly and substituting for the Green's function $\mathbf{G}^{(n+1)}_{n+1,j}$ using (\ref{FiveDA}) gives (\ref{FiveBA}).

\end{document}